\documentclass{mn2e} 
\usepackage{graphics}
\usepackage{epsfig}

\def\ni{\noindent}                                       
\def\etal{et\thinspace al.\ }                               
 
\title[An atlas of Calcium triplet spectra of active galaxies]
       {An atlas of Calcium triplet spectra of active galaxies}

\author[Garcia-Rissmann \etal] 
       {A. Garcia-Rissmann$^{1}$\thanks{E-mail: aurea@astro.ufsc.br},
	 L. R. Vega$^{1,2}$\thanks{E-mail: luis@astro.ufsc.br},
	 N. V. Asari$^{1}$\thanks{E-mail: natalia@astro.ufsc.br}, 
	 R. Cid Fernandes$^{1}$\thanks{E-mail: cid@astro.ufsc.br},
	 \newauthor
	 H. Schmitt$^{3,4}$\thanks{E-mail: hschmitt@css.nrl.navy.mil},
	 R. M. Gonz\'alez Delgado$^{5}$\thanks{E-mail: rosa@iaa.es},
	 T. Storchi-Bergmann$^{6}$\thanks{thaisa@if.ufrgs.br}\\
	   $^{1}$ Depto.\ de F\'{\i}sica - CFM - Universidade Federal de
	   Santa Catarina, C.P. 476, 88040-900, Florian\'opolis, SC, Brazil\\
	   $^{2}$ Observatorio Astron\'omico de C\'ordoba, Laprida 854, 5000, 
	   C\'ordoba, Argentina\\
	   $^{3}$ Remote Sensing Division, Code 7210, Naval Research Laboratory,
4555 Overlook Avenue, SW, Washington, DC 20375\\
	   $^{4}$ Interferometric Inc., 14120 Parke Long Court, 103, Chantilly, VA20151\\
	   $^{5}$ Instituto de Astrof\'{\i}sica de Andaluc\'{\i}a (CSIC),
	   P.O. Box 3004, 18080 Granada, Spain\\
	   $^{6}$ Instituto de F\'{\i}sica, Universidade Federal do Rio
	   Grande do Sul, C.P. 15001, 91501-970, Porto Alegre, RS, Brazil}

\begin{document}

\maketitle

\begin{abstract} 

We present a spectroscopic atlas of active galactic nuclei covering
the region around the $\lambda\lambda$8498, 8542, 8662 Calcium triplet
(CaT). The sample comprises 78 objects, divided into 43 Seyfert 2s, 26
Seyfert 1s, 3 Starburst and 6 normal galaxies.  The spectra pertain to
the inner $\sim 300$ pc in radius, and thus sample the central
kinematics and stellar populations of active galaxies. The data are
used to measure stellar velocity dispersions ($\sigma_\star$) both
with cross-correlation and direct fitting methods. These measurements
are found to be in good agreement with each-other and with those in
previous studies for objects in common. The CaT equivalent width is
also measured.  We find average values and sample dispersions of
$W_{\rm CaT}$ of $4.6\pm2.0$, $7.0\pm$ and $7.7\pm$1.0 \AA\ for
Seyfert 1s, Seyfert 2s and normal galaxies, respectively.  We further
present an atlas of [SIII]$\lambda$9069 emission line profiles for a
subset of 40 galaxies. These data are analyzed in a companion paper
which addresses the connection between stellar and Narrow Line Region
kinematics, the behaviour of the CaT equivalent width as a function of
$\sigma_\star$, activity type and stellar population properties.
\end{abstract}


\section{Introduction}

\label{sec:Introduction}

Fifteen years ago, Terlevich, D\'{\i}az \& Terlevich (1990,
hereinafter TDT) carried out the first systematic study of the
$\lambda\lambda$8498,8542,8662 \AA\ absorption lines of the Ca II ion
in active galactic nuclei (AGN). The main focus of that pioneer work
on the ``Calcium Triplet'' (CaT) was on the equivalent width of this
feature ($W_{\rm CaT}$), which provides both a stellar population
diagnostic and a tool to investigate the presence of an underlying
non-stellar continuum.  Most of the AGN in the TDT sample were type 2
Seyferts.  Their main finding was that $W_{\rm CaT}$ is remarkably
similar for Seyfert 2s and normal galaxies, implying that the
non-stellar featureless continuum (FC) invoked to account for the
dilution of optical absorption lines in these objects either is not
featureless at all or somehow disappears between optical and
near-infrared (NIR) wavelengths. The interpretation advanced by TDT
was that both the optical FC and the CaT lines are produced by a
nuclear starburst. Subsequent work by the same group suggests that
this may also apply to at least some type 1 Seyferts (Jim\'enez-Benito
\etal 2000), although in the more active of these objects the AGN
continuum does cause some dilution of the CaT.

Since then, the existence of starbursts around Seyfert 2 nuclei has
been established by both indirect means (Cid Fernandes \& Terlevich
1995; Heckman \etal 1995; Oliva \etal 1999) and direct detections of
young stars by optical--UV imaging and spectroscopy (Heckman \etal
1997; Gonz\'alez Delgado, Heckman \& Leitherer 2001).  While these
studies confirmed that the optical FC of Seyfert 2s is predominantly
due to recent star-formation, there are still doubts as to whether the
CaT is produced by these starbursts or by older stars in the host
galaxy bulge. In other words, it is not clear whether the CaT can be
used to diagnose the presence of starbursts.  Indeed, the mere fact
that $W_{\rm CaT}$ shows little variation among galaxies of widely
different morphological and spectroscopic properties indicates that
this feature may not be as simple a tracer of stellar populations as
initially thought.  Recent empirical and theoretical work reinforce
this idea (Cenarro \etal 2003, 2004; Saglia \etal 2002;
Falc\'on-Barroso \etal 2003; Michielsen \etal 2003; Thomas, Maraston
\& Bender 2003; Vazdekis \etal 2003), and show that there is still
much to be learned about the CaT behaviour even in normal galaxies.

A more widespread use of the CaT nowadays is to measure stellar
velocity dispersions ($\sigma_\star$). This was the approach followed
by Nelson \& Whittle (1995 and 1996; hereinafter collectively revered
to as NW) in their comprehensive study of stellar and gaseous
kinematics of AGN.  The discovery of the relation between black-hole
mass ($M_{\rm BH}$) and $\sigma_\star$ (Ferrarese \& Merrit 2000;
Gebhardt \etal 2000; Tremaine \etal 2002) brought renewed interest in
this kind of work. Indeed, most of the current observational CaT
studies in AGN are geared towards using this spectroscopic feature as
an indirect black-hole weighing-scale (Ferrarese \etal 2001; Barth, Ho
\& Sargent 2002, 2003; Filippenko \& Ho 2003; Barth \etal 2004; Botte
\etal 2004; Nelson \etal 2004; Onken \etal 2004; Barth, Greene \& Ho
2005). Finally, velocity dispersions are also useful to investigate
stellar populations. Combined with size and luminosity measurements,
$\sigma_\star$ allows the estimation of the mass-to-light ratio, which
is a strong function of the age in stellar systems. There have been
few applications of this idea to AGN, but the results reported so far
seem to fit the scenario where active nuclei tend to be surrounded by
stellar populations younger than those typical of elliptical galaxies
and bulges (NW; Oliva \etal 1995; 1999).

This brief summary illustrates that there is plenty of motivation to
study the CaT in both active and normal galaxies. In this paper we
present an atlas of CaT spectra and related data products for a sample
of 78 galaxies, most of which have active nuclei. This material is
used in a companion paper (Vega \etal 2005, hereafter Paper II) to
address issues such as the connection between nebular and stellar
kinematics, the sensitivity of $W_{\rm CaT}$ to stellar population
properties, and constraints on the contribution of a non-stellar
component to the NIR spectra of AGN.

This paper is organized as follows. Section \ref{sec:data} describes
the sample, observations and data reduction. Section \ref{sec:atlas}
presents our atlas of CaT spectra, as well as [SIII]$\lambda$9069
emission line profiles for a subset of the objects.  Measurements of
stellar velocity dispersions are presented Section \ref{sec:vel_disp},
while Section \ref{sec:EWs} presents results on the equivalent width
of the CaT. Finally, Section \ref{sec:Conclusions} summarizes our main
results.

\section{Observations}

\label{sec:data}

The data presented here were obtained in 6 runs in 3 different
telescopes: two at the 1.52m ESO-La Silla (39 galaxies), two at the
KPNO 2.1m (25 galaxies), and two at the KPNO 4m telescope (16
galaxies). Although the original projects had somewhat different
specific goals, they all centered on the measurement of the CaT in
AGN. We have thus decided to merge all the CaT-related data in a
single atlas containing the nuclear spectra and associated
data-products, processed in a way as homogeneous as possible. This
section describes the observations, reduction process and the general
sample properties.

\subsection{ESO 1.52m observations}

\label{sec:ESO_observations}

\begin{table*}
\begin{center}
\begin{tabular}{lccccccl}
\multicolumn{8}{c}{} \\
\hline
  Object   & Activity    & $v_{rad}$ (km/s)& Morph Type & T      &  Date        & Exp.\ time (s)&  Flag \\ \hline
  NGC 526A & Sy1.5       & 5725       & S0 pec?		& 0.0	 &  04 Oct 2002 & (2x)1800	&  c    \\	
  NGC 526B &  Normal     & 5669       & SB0: pec	& 0.0	 &  04 Oct 2002 & (1x)1500	&  c    \\	
  NGC 1125 &   Sy2       & 3277       & SB(r)0+		& -1.0   &  30 Sep 2002 & (3x)1500	&  c    \\	
  NGC 1140 & HII/Sy2     & 1501       & IBm pec		& 10.0   &  03 Oct 2002 & (3x)1500 	&  c    \\	
  NGC 1365 &   Sy1.8     & 1636       & (R')SBb(s)b	& 3.0    &  02 Oct 2002 & (3x)1500 	&  d    \\  
  NGC 1380 &  Normal     & 1877       & SA0		& 0.0	 &  30 Sep 2002 & (2x)1500	&  c    \\	
  NGC 1433 &   Sy2       & 1075       & (R'$_1$)SB(rs)ab& 2.0	 &  01 Oct 2002 & (3x)1500 	&  b    \\
  NGC 1672 &   Sy2       & 1350       & (R'$_1$)SB(r)bc & 4.0 	 &  03 Oct 2002 & (2x)1500 	&  a    \\
  NGC 1808 &   Sy2       & 1000       & (R'$_1$)SAB(s:)b& 3.0	 &  04 Oct 2002 & (2x)1200	&  a    \\
  NGC 2997 & Normal      & 1087       &	SA(s)c		& 5.0	 &  11 Mar 2002 & (2x)1200 	&  a    \\   
  NGC 3081 &   Sy2       & 2385       & (R'$_1$)SAB(r)0 & 0.0	 &  09 Mar 2002 & (4x)1200 	&  a    \\ 
  NGC 3115 &   Sy2       &  720       &	S0-		& -3.0	 &  11 Mar 2002	& (1x)600 	&  a    \\  
  NGC 3256 &   HII       & 2738       & Pec; merger     & ?	 &  12 Mar 2002	& (1x)900,(1x)1200 & b  \\	
  NGC 3281 &   Sy2       & 3200       & SAB(rs+)a       & 1.0	 &  10 Mar 2002 & (4x)1200 	&  b    \\   
  NGC 3783 &   Sy1       & 2717       & (R')SB(r)a      & 1.0	 &  09 Mar 2002 & (5x)1200 	&  a    \\	
  NGC 4507 &   Sy1.9     & 3538       & SAB(s)ab        & 2.0	 &  11 Mar 2002 & (3x)1800 	&  a    \\   
  NGC 4593 &   Sy1       & 2698       & (R)SB(rs)b      & 3.0	 &  11 Mar 2002 & (3x)1800 	&  c    \\   
  NGC 4748 &   Sy1 (NLSy1) & 4386     & Sa		& 1.0*	 &  12 Mar 2002 & (3x)1800 	&  c    \\   
  NGC 4968 &   Sy2       & 2957       &(R')SAB0$^0$     & -2.0	 &  12 Mar 2002 & (2x)1800 	&  a    \\	
  NGC 5135 &   Sy2       & 4112       & SB(1)ab	        & 2.0	 &  11 Mar 2002 & (2x)1800 	&  a    \\   
  NGC 6300 &   Sy2       & 1110       & SB(rs)b         & 3.0	 &  12 Mar 2002 & (2x)1800 	&  a    \\	
  NGC 6814 &  Sy1.5      & 1563       & SAB(rs)bc       & 4.0	 &  04 Oct 2002 & (3x)1500	&  b    \\   
  NGC 6860 &   Sy1       & 4462       & (R')SB(r)ab     & 2.0	 &  03 Oct 2002 & (3x)1500	&  c    \\   
  NGC 6907 &  Normal     & 3161       & SB(s)bc         & 4.0	 &  30 Set 2002 & (3x)1500	&  c    \\   
  NGG 7130 &  Sy2/L      & 4842       & Sa pec          & 1.0	 &  30 Sep 2002 & (3x)1500	&  b    \\   
  NGC 7172 &   Sy2       & 2603       & Sa pec sp       & 1.0	 &  02 Oct 2002 & (3x)1500	&  a    \\   
  NGC 7184 &  Normal     & 2617       & SB(r)c          & 5.0	 &  02 Oct 2002 & (3x)1500	&  a    \\   
  NGC 7410 &  Sy2/L      & 1751       & SB(s)a          & 1.0	 &  04 Oct 2002 & (2x)1800	&  a    \\   
  NGC 7496 &   Sy2       & 1649       & (R':)SB(rs)bc   & 4.0	 &  02 Oct 2002 & (3x)1500	&  c    \\   
  NGC 7582 &   Sy2       & 1575       &(R'$_1$)SB(s)ab  & 2.0	 &  03 Oct 2002 & (2x)1800	&  a    \\	
  NGC 7590 &   Sy2       & 1596       & S(r?)bc	        & 4.0	 &  30 Sep 2002 & (3x)1500	&  a    \\   
  NGC 7714 &  HII/L      & 2798       & SB(s)b:pec	& 3.0	 &  04 Oct 2002 & (2x)1200,(1x)1500 & b \\	
  IC  2560 &   Sy2       & 2925       & (R':)SB(r)bc    & 4.0    &  12 Mar 2002	& (3x)1800 	&  a    \\   
  IC  3639 &   Sy2       & 3275       & SB(rs)bc:       & 4.0	 &  10 Mar 2002 & (3x)1800 	&  a    \\   
  IC  5169 &   Sy2       & 3016       &(R'$_1$)SAB(r)0+ & -1.0	 &  01 Oct 2002 & (3x)1500	&  c    \\   
  ESO 362G08 & Sy2       & 4785       & Sa 	        & 1.0*	 &  09 Mar 2002 & (4x)1200 	&  a    \\   
  ESO 362G18 & Sy1.5     & 3790       & S0/a	        & 0.0*	 &  10 Mar 2002 & (3x)1800 	&  a    \\   
  MCG-6.30.15 &Sy1.2     & 2323       & E-S0	        & -2.0*	 &  10 Mar 2002 & (3x)1800 	&  a    \\
  Mrk 1210 &   Sy2       & 4046       &	Sa	        & 1.0*	 &  11 Mar 2002 & (3x)1800  	&  b    \\ \hline
\end{tabular}
\end{center}
\caption{Log of the ESO 1.52m telescope observations. Columns 2,3 and
4 list the activity type, radial velocity (in km/s) and Hubble class,
all extracted from NED.  Column 5 lists the numerical Hubble type,
taken from RC3 (de Vaucouleurs \etal 1991), (except for those marked
with asterisks, unavailable in the catalog, 
and whose T-types were attributed by us based on the Hubble morphological type).  Column 8 lists a quality flag (see Section
\ref{sec:CaT_atlas}).}
\label{tab:log_ESO}
\end{table*}

Most of the Southern objects in the sample have been observed with the
Boller \& Chivens spectrograph coupled at the Cassegrain focus
(f/14.9) of the now extinct 1.52m telescope, located in ESO-La Silla
(Chile), during two runs (2002 March and October). Similar setups have
been adopted in both runs, with a grating of 900 l/mm (\# 5) centered
at about 7230 \AA, giving a dispersion of 1.32 \AA~pix$^{-1}$ on the CCD
\#38 (2688$\times$512 pixels$^2$, each pixel with a 15$\mu$m
size). The slit width of 2$^{\prime\prime}$ adopted for all the program objects
(which comprise the galaxies and the template stars) provided a
resolution $\sigma_{\rm inst}$ of about 56 and 44 km/s for the
spectra of March and October, respectively, as measured through sky
emission lines. The slit was always aligned in the E-W direction
(P.A. $= 90^\circ$), and was long enough (4.5$^{\prime}$) as to guarantee
the inclusion of enough sky to allow its subtraction from the galaxies
spectra.  The plate scale on the CCD was 0.82$^{\prime\prime}$ pix$^{-1}$, and the
wavelengths covered by the observations ranged from $\sim$ 6300 to
9500 \AA.  Wavelength calibration was performed using a HeNeArFe lamp
spectra taken in each telescope position.  A log of the observations
is presented in Table \ref{tab:log_ESO}. A series of velocity standard
stars, listed in Table \ref{tab:VelocityStandards}, have also been
observed with the same setup and used as templates in the
determination of $\sigma_\star$ (\S\ref{sec:vel_disp}).

Since fringing effects in the NIR can be a serious concern (see
below), internal flats made with a Quartz lamp have also been acquired
(only in the first run) for every telescope pointing. Twilight flats
have also been taken, generally at the sunset, to be used for the
illumination correction. Spectrophotometric standard stars from Hamuy
\etal (1994), Oke (1990), Massey \etal (1988) and Massey \& Gronwal
(1990) were observed (at least 2 per night), always with the slit wide
open (5--8$^{\prime\prime}$), for flux calibration. Given the variable
atmospheric conditions (particularly in the October run) our absolute
flux scale is uncertain, but this has no consequence for the results
reported in this paper, since we report only relative measurements.

\subsection{KPNO 2.1m observations}

\label{sec:KPNO_observations}

\begin{table*}
\begin{center}
\begin{tabular}{lccccccl}
  \multicolumn{8}{c}{} \\
\hline
  Object        & Activity      & $v_{rad}$ (km/s) & Morph Type  & T &  Date       & Exp.\ time (s) &  Flag \\ \hline
 Mrk 40         & Sy1          	& 6323	& S0 pec	& -2.0   & 20 Feb 2003 & (4x)1200 &  b \\
 Mrk 79         & Sy1.2        	& 6652	& SBb   	&  3.0   & 14 Nov 2002 & (3x)1200 &  d \\
 Mrk 372        & Sy1.5        	& 9300	& S0/a  	&  0.0*  & 20 Feb 2003 & (3x)1200 &  a \\
 Mrk 461        & Sy2          	& 4856	& S0    	& -2.0*  & 19 Feb 2003 & (3x)1200 &  b \\
 Mrk 516        & Sy1.8        	& 8519	& Sc    	&  6.0*  & 15 Nov 2002 & (3x)1200 &  b \\
 Mrk 705        & Sy1.2        	& 8739	& S0?   	& -2.0   & 15 Nov 2002 & (3x)1200 &  c \\
 Mrk 915        & Sy1  	       	& 7228	& Sb    	&  3.0*  & 15 Nov 2002 & (3x)1200 &  b \\
 Mrk 1210       & Sy1/Sy2      	& 4046	& Sa    	&  1.0*  & 13 Nov 2002 & (3x)1200 &  a \\
 Mrk 1239       & Sy1.5 (NLSy1) & 5974  & E-S0  	& -3.0*  & 19 Feb 2003 & (3x)1200 &  d \\
 UGC 3478       & Sy1.2        	& 3828	& Sb    	&  3.0   & 20 Feb 2003 & (4x)1200 &  d \\
 UGC 1395       & Sy1.9        	& 5208	& SA(rs)b	&  3.0   & 14 Nov 2002 & (3x)1200 &  b \\
 UGC 12138      & Sy1.8        	& 7487	& SBa   	&  1.0   & 13 Nov 2002 & (3x)1200 &  b \\
 UGC 12348      & Sy2          	& 7631	& Sa    	&  1.0   & 15 Nov 2002 & (3x)1200 &  a \\
 NGC 1019       & Sy1          	& 7251	& SB(rs)bc  	&  4.0   & 14 Nov 2002 & (3x)1200 &  a \\
 NGC 1142       & Sy2  	       	& 8648	& S pec (Ring B)&  1.0   & 14 Nov 2002 & (3x)1200 &  c \\
 NGC 1241       & Sy2          	& 4052	& SB(rs)b   	&  3.0   & 15 Nov 2002 & (3x)1200 &  a \\
 NGC 2639       & Sy1.9        	& 3336	& (R)SA(r)a:?   &  1.0   & 13 Nov 2002 & (3x)1200 &  a \\
 NGC 6951       & L/Sy2        	& 1424	& SAB(rs)bc	&  4.0   & 14 Nov 2002 & (2x)1200 &  a \\
 NGC 7469       & Sy1.2        	& 4892	& (R')SAB(rs)a  &  1.0   & 13 Nov 2002 & (3x)1200 &  a \\
 IRAS 01475-0740& Sy2          	& 5296	& E-S0  	& -3.0*  & 13 Nov 2002 & (3x)1200 &  c \\
 IRAS 04502-0317& Sy2          	& 4737	& SB0/a 	&  0.0   & 15 Nov 2002 & (3x)1200 &  c \\
 MCG -01-24-012 & Sy2          	& 5936	& SAB(rs)c: 	&  5.0   & 14 Nov 2002 & (3x)1200 &  c \\
 MCG -02-08-039 & Sy2         	& 8989	& SAB(rs)a pec:	&  1.0   & 15 Nov 2002 & (3x)1200 &  b \\
 MCG +8-11-11   & Sy1.5        	& 6141	& SB0   	& -2.0*  & 13 Nov 2002 & (3x)1200 &  d \\
 Akn 564        & Sy1.8 (NLSy1) & 7400  & SB    	&  0.0	 & 14 Nov 2002 & (3x)1200 &  d \\
\hline
\end{tabular}
\end{center}
\caption{Log of the KPNO 2.1m telescope observations.}
\label{tab:log_KPNO}
\end{table*}

A total of 25 Northern galaxies were observed on 2 observing runs with
the KPNO 2.1m telescope, on the nights of 2002-Nov-12/13-14/15 and
2003-Feb-17/18-19/20. The observational setup was the same in both
runs, resulting in spectra of similar quality to those obtained for
the ESO sample.  We used the Gold Camera Spectrograph with grating
\#35 and a slit width of 2$^{\prime\prime}$. The slit was oriented in
the E-W direction during the first run and in the N-S direction in the
second one. This configuration gives a plate scale of
0.78$^{\prime\prime}$ pix$^{-1}$, a spectral resolution of 1.24
\AA~pix$^{-1}$ and a wavelength coverage of $\sim 6800$--9300 \AA.  The
spectral resolution for these data is $\sigma_{\rm inst} \sim 57$
km/s. A log of the observations is presented in
Table~\ref{tab:log_KPNO}.
 
The observation of each galaxy was preceded and followed by internal
Quartz lamp flat-field frames (for fringing corrections) and HeNeAr
wavelength calibration frames.  In the case of flux and velocity
standard stars, which required only short integrations, the fringe
pattern does not vary significantly, so we obtained spectra of the
Quartz and HeNeAr lamps only once, either before or after the
observation of the star.  During the first observing run (November) we
obtained a series of twilight flats, which were used for illumination
correction. However, due to bad weather conditions, we were not able
to obtain twilight flats during the February run, having to resort to
a combination of the program frames (excluding the regions of the
spatial profiles) for the illumination correction.
 
Throughout the night we observed a series of velocity standard stars
of various spectral types (Table \ref{tab:VelocityStandards}), using
the same slit width used for the observation of the galaxies.
Finally, at least 3 spectrophotometric standard stars were observed
every night, using a 5$^{\prime\prime}$ slit.  We observed
BD$+$17~4708, G191B2B and Feige~34 in the November run and Feige~66,
HZ~44, Hiltner~600 and Feige~34 in the February run. As was the case
in the ESO observations, not all of the KPNO observations were done
under photometric conditions, however, this does not affect the
outcome of this project.

\subsection{KPNO 4m observations}

\label{sec:ROSA_observations}

\begin{table*}
\begin{center}
\begin{tabular}{lccccccl}
\multicolumn{8}{c}{} \\
\hline
  Object   & Activity& $v_{rad}$ (km/s)& Morph Type & T &  Date       & Exp.\ time (s) &  Flag \\ \hline
  Mrk 0001 &   Sy2   & 4780       & S?		&   1.0	&  11 Oct 1996 & (2x)1800  	&  a    \\
  Mrk 0003 &   Sy2   & 4050       & S0:		&  -1.0	&  15 Feb 1996 & (2x)1800  	&  b    \\ 
  Mrk 0078 &   Sy2   & 11137      & SB		&  ?	&  15 Feb 1996 & (2x)1800  	&  b    \\	
  Mrk 0273 & Sy2/L   & 11326      & Ring Gal pec&  ?	&  15 Feb 1996 & (3x)1800  	&  c    \\   
  Mrk 0348 &   Sy2   & 4507       & SA(s)0/a	&   1.0 &  11 Oct 1996 & (2x)1800  	&  a    \\	
  Mrk 0573 &   Sy2   & 5174       & SAB(rs)0+   &  -1.0	&  11 Oct 1996 & (2x)1800  	&  a    \\   
  Mrk 1066 &   Sy2   & 3605       & SB(s)0+	&  -1.0	&  11 Oct 1996 & (1x)1800,(1x)900 & a   \\   
  Mrk 1073 &   Sy2   & 6998       & SB(s)b	&   3.0	&  11 Oct 1996 & (1x)1800,(1x)900 & a   \\   
  NGC 0205 & Normal  & -241       & dE          &  -5.0 &  15 Feb 1996 & (1x)300,(2x)600  & a   \\   
  NGC 1068 & Sy1/2   & 1137       & SA(rs)b	&   3.0	&  11 Oct 1996 & (2x)900 	&  a    \\ 
  NGC 1386 &   Sy2   &  868       & SB(s)0+	&  -1.0	&  11 Oct 1996 & (2x)1200  	&  a    \\	
  NGC 2110 &   Sy2   & 2335       & SAB0-	&  -3.0	&  15 Feb 1996 & (1x)1800  	&  a    \\	
  NGC 4339 &   Sy2   & 1289       & E           &  -5.0 &  15 Feb 1996 & (1x)300,(2x)600 & a    \\   
  NGC 5929 &   Sy2   & 2492       & Sab:pec	&   2.0	&  15 Feb 1996 & (1x)1800,(1x)900 & a   \\	
  NGC 7130 & Sy2/L   & 4842       & Sa pec	&   1.0	&  11 Oct 1996 & (1x)1800  	& b     \\   
  NGC 7212 &   Sy2   & 7984       & S		&  ?	&  11 Oct 1996 & (2x)1800  	& a     \\
\hline
\end{tabular}
\end{center}
\caption{Log of the KPNO 4m Mayall telescope observations.}
\label{tab:log_ROSA}
\end{table*}

Finally, we have incorporated to our data set the CaT observations of
14 Seyfert 2s and 2 normal galaxies taken in the KPNO 4m Mayall
telescope during two runs in 1996.  Table \ref{tab:log_ROSA} describes
these observations.  The spectra were taken with a dispersion of 1.52
\AA~pix$^{-1}$ covering the unvignetted spectral ranges 6600--9100 \AA\
(February 1996 run) and 7400--9800 \AA\ (October 1996 run). The slit
width of 1.5$^{\prime\prime}$ was set at the parallactic angle. These data were
partly described in Gonz\'alez Delgado \etal (2001), to which we refer
the reader for details of the observations and reduction process.

\begin{table}
\begin{center}
\begin{tabular}{lcl}
\hline
Star      & Spectral Type & Telescope \\ \hline
HD   9737 & F0III & KPNO 2.1m  \\
HD   9748 & K0III & KPNO 2.1m  \\
HD  19136 & K0III & KPNO 2.1m  \\
HD  21910 & K0III & KPNO 4m    \\
HD  23962 & K5III & KPNO 2.1m  \\
HD  31805 & F0III & KPNO 2.1m  \\
HD  39008 & K3III & KPNO 2.1m  \\
HD  39833 & G0III & KPNO 2.1m  \\
HD  41589 & K0III & KPNO 2.1m  \\
HD  62564 & K0III & KPNO 2.1m  \\
HD  71597 & K2III & KPNO 4m    \\
HD  77189 & K5III & KPNO 2.1m  \\
HD  84059 & F0III & KPNO 2.1m  \\
HD  87018 & K3III & ESO 1.52m  \\
HD  89885 & K0III & ESO 1.52m  \\
HD 113678 & K0III & ESO 1.52m  \\
HD 116535 & K0III & ESO 1.52m  \\
HD 116565 & K0III & ESO 1.52m  \\
HD 119171 & K0III & KPNO 2.1m  \\
HD 120572 & K3III & ESO 1.52m  \\
HD 121138 & K0III & ESO 1.52m  \\
HD 121883 & K0III & ESO 1.52m  \\
HD 122665 & K5III & ESO 1.52m  \\
HD 124990 & K0III & ESO 1.52m  \\
HD 127740 & F5III & KPNO 2.1m  \\
HD 128529 & K5III & ESO 1.52m  \\
HD 132151 & K0III & ESO 1.52m  \\
HD 139447 & K5III & ESO 1.52m  \\
HD 143393 & K2III & KPNO 4m    \\
HD 143976 & K5III & ESO 1.52m  \\
HD 151817 & K3III & ESO 1.52m  \\
HD 160413 & K3III & ESO 1.52m  \\
HD 195527 & K0III & KPNO 2.1m  \\
HD 209543 & K0III & KPNO 2.1m  \\
HD 219656 & K0III & KPNO 2.1m  \\
HD 258403 & F0III & KPNO 2.1m  \\ \hline
\end{tabular}
\end{center}
\caption{Velocity standard stars.}
\label{tab:VelocityStandards}
\end{table}

\subsection{Reduction}

\label{sec:Reduction}

Two major problems affect observations and data reductions in the NIR
region: the significant contamination by atmospheric emission lines
and fringing. In this section we describe how we have dealt with these
problems in the reduction of the ESO and KPNO 2.1m data.  

Fringing is caused by the back and forth scattering of the NIR light
in coated CCDs. Here we adopted a careful procedure to eliminate
fringes, or, at least, to minimize their effects. Instead of using
dome-flats, we flat-fielded our data using Quartz lamp spectra, since
fringes (present in all images) tend to change their locations as the
telescope moves. Although for every telescope position one lamp flat
was taken, in the first ESO run we decided to search for those flats
which reduced the fringing patterns in the final spectrum, following
the recipe given in Plait \& Bohlin (1997) for the STIS observations
(where a library of fringing patterns is available for observers). In
order to determine the lamp flat which minimized this effect, we
performed the reduction and extraction process using each of them, for
every object spectrum; the extracted spectra were then normalized in
the region of interest by smoothed versions of themselves (using a 50
pixels boxcar filter), and finally an auto-correlation analysis over
the normalized extracted spectrum was made. For spectra where fringing
is more conspicuous, the rms in this auto-correlation function reaches
higher values.  By selecting the spectra whose auto-correlation
function had the lowest rms in the CaT region, we automatically
selected those with the smallest fringing patterns. This procedure has
shown to slightly improve the final spectra, compared to no fringing
correction. In any case, we have verified that these corrections have
little effect upon the main data products reported in this paper,
namely, stellar velocity dispersions and the CaT equivalent width.

The subtraction of atmospheric lines can introduce spurious spikes in
the extracted spectra, specially near the most intense ones. This is
caused by small fluctuations in the width and location of such lines
along the spatial axis, so a sky region as close as possible to the
extraction window is preferred. Since this often cannot be the case in
observations of extended sources, we adopted a careful procedure for
background removal. Possible slightly misalignments of the dispersion
axis along the CCD lines were corrected applying IRAF's task {\it
identify} to all 2D images, and a bi-dimensional function of order $6
\times 6$ was fitted through {\it fitcoords}.  Before the fit, we
analyzed carefully each object's spatial profile, to determine which
CCD regions were at (or as close as possible to) the sky level; we
only included in the fit the CCD regions of interest (extraction
window + background), in order to achieve a better precision
(residuals $\leq$ 0.2 \AA~pix$^{-1}$).  Finally, we used the task {\it
transform} to apply the wavelength correction to all the images before
the spectral extraction.  Headers from the IHAP/ESO acquisition system
have been completed with the task {\it asthedit} from IRAF.

For both the ESO and KPNO 2.1m data we have extracted nuclear spectra
adding the central 3 pixels (through the optimal extraction algorithm,
Horne 1986).  This corresponds to a spatial scale of 2.46$^{\prime\prime}$ and 
2.34$^{\prime\prime}$ for the ESO and KPNO 2.1m objects, respectively.

The correction for atmospheric extinction was applied using the
specific observatories data. Galactic reddening was corrected using
the Cardelli, Clayton \& Mathis (1989) law with $R_V = 3.1$ and the
$A_B$ values from Schlegel, Finkbeiner \& Davis (1998) as listed in
NED. Atmospheric telluric lines (due mainly to H$_2$O and O$_2$) have
not been corrected. Nevertheless, as noticed by NW, these affect the
measurement of the CaT only in galaxies with redshifts greater than
about 8000 km/s, i.e., in only 4 out of 78 objects of our sample.

\subsection{Sample properties}

\label{sec:Sample}

\begin{figure*}
\includegraphics[bb= 50 500 610 710,width=17cm]{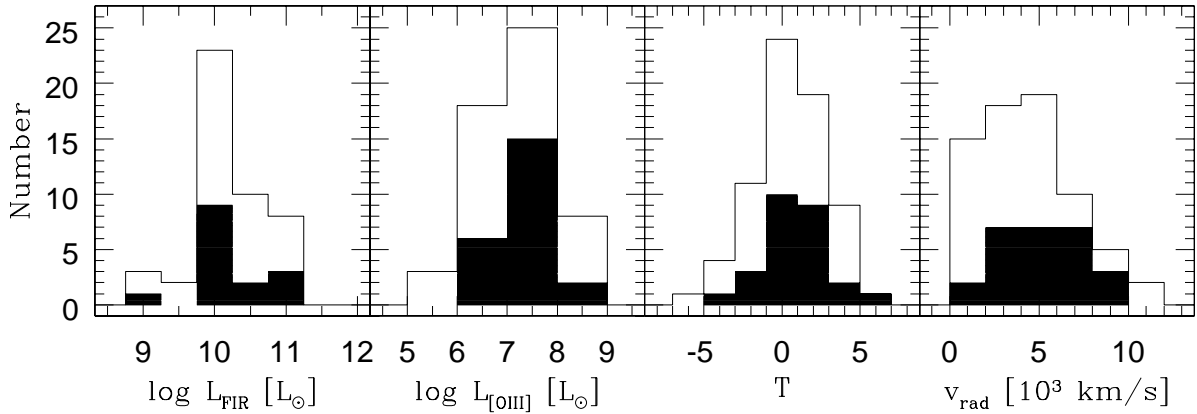}
\caption{Sample properties. Empty areas correspond to Seyfert 2s and
filled boxes to Seyfert 1s.}
\label{fig:sample}
\end{figure*}

\begin{table*}
\small
\begin{centering}
\begin{tabular}{lrrrrrrrrr}
\multicolumn{10}{c}{Summary of Sample Properties} \\ \hline
                            &
\multicolumn{3}{c}{Seyfert 2s}  &
\multicolumn{3}{c}{Seyfert 1s}  &
\multicolumn{3}{c}{Non AGN} \\ 
               & 
N              & 
range          & 
mean $\pm$ rms &
N              & 
range          & 
mean $\pm$ rms &
N              & 
range          & 
mean $\pm$ rms 	\\ \hline
T-type                                 &  43   &  (-5,5)     &  1.0 $\pm$ 2.3 
                                       &  26   &  (-3,6)     &  1.4 $\pm$ 2.1 
                                       &   9   &  (-5,10)    &  2.6 $\pm$ 4.2 \\
distance [Mpc]                         &  43   &  (10,151)   &   52 $\pm$ 36  
                                       &  26   &  (21,124)   &   69 $\pm$ 29  
                                       &   9   &  (1,76)     &   32 $\pm$ 21  \\
$r_{ap}$ [pc]                          &  43   &  (54,715)   &  285$\pm$187  
                                       &  26   &  (126,734)  &  414$\pm$172  
                                       &   9   &  (3,459)    &  193$\pm$128   \\
$\log L_{\rm [OIII]}$ [L$_\odot$]      &  31   &  (5.6,8.7)  &   7.2$\pm$ 0.8 
                                       &  23   &  (6.2,8.1)  &   7.4$\pm$ 0.6 
                                       &       &             &                \\
$\log L_{\rm FIR}$ [L$_\odot$]         &  31   &  (8.9,11.1) &  10.2$\pm$ 0.5 
                                       &  15   &  (9.2,11.2) &  10.1$\pm$ 0.5 
                                       &   7   &  (6.0,11.3) &   9.6$\pm$ 1.7 \\ \hline
\end{tabular}
\end{centering}
\caption{Statistics of selected properties of the sample, divided into
Sey 2s, Sey 1s and other galaxies. For each property we list the
range, mean value and sample dispersion, and the number of objects for
which the corresponding quantity was available.  Luminosities were
computed for $H_0 = 75$ km$\,$s$^{-1}$Mpc$^{-1}$. All properties were
compiled from NED and the literature.}
\label{tab:sample}
\end{table*}

Tables \ref{tab:log_ESO}--\ref{tab:log_ROSA} list some properties of
the galaxies in the sample: including morphology, recession velocity
and activity class, all extracted from the NED database. Since this
list results from the merging of different observational programs,
none of which aimed completeness in any sense, the resulting sample is
a rather mixed bag of objects. Overall, however, this is a
representative sample of Seyfert galaxies in the nearby universe.

In total, we have 80 spectra of 78 galaxies, divided into 43 Seyfert
2s, 26 Seyfert 1s (including intermediate Seyfert types and Narrow
Line Seyfert 1s) and 9 non-active galaxies, including 3 Starburst
nuclei. Mrk 1210 and NGC 7130 were observed twice with different
telescopes. Though we keep only the better spectra in the atlas, these
duplicate measurements are useful to check uncertainties in our
measurements (Section \ref{sec:DFM}). Fig \ref{fig:sample} and Table
\ref{tab:sample} summarize some statistics of the sample, divided into
Seyfert 1s, Seyfert 2s and non-active galaxies.  The Far IR fluxes
were compiled from NED.  [OIII]$\lambda$5007 fluxes were compiled from
a number of papers (Whittle 1992; Bassani \etal 1999; Schmitt \etal
2003; Storchi-Bergmann, Kinney \& Challis 1995). In case of duplicate data, we
favour measurements obtained under large extractions (to include more
of the Narrow Line Region emission) and corrected by
reddening. Inevitably, the resulting [OIII] luminosities are very
inhomogeneous, and should be regarded as uncertain by a factor of $\sim
2$.  The table shows that we span a wide range of morphological types,
distances, [OIII]$\lambda$5007 and Far-IR luminosities.

\section{The Atlas}

\label{sec:atlas}

\begin{figure*}
\includegraphics[width=18cm]{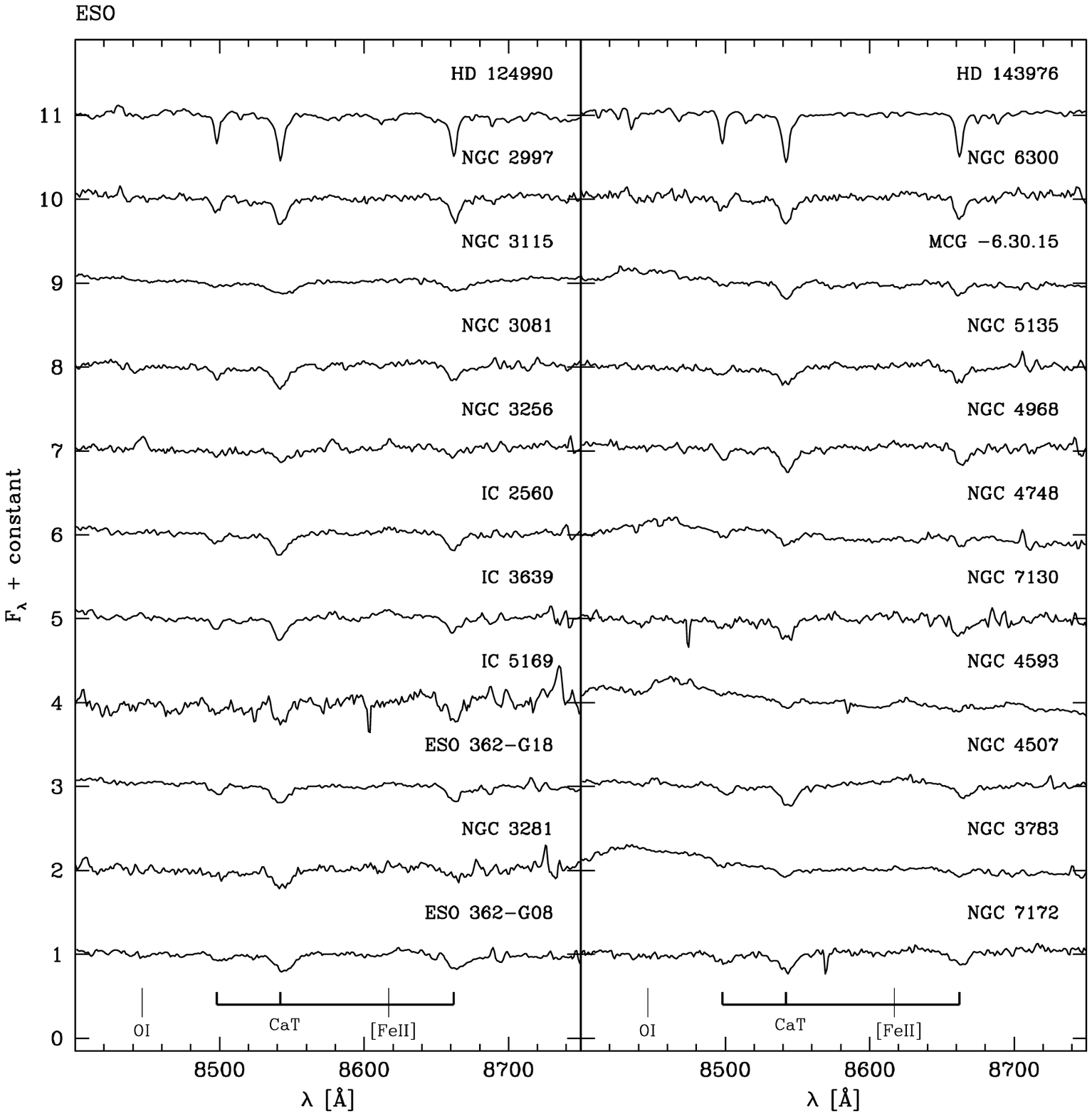}
\caption{CaT spectra for the ESO 1.52m observations. All spectra are
normalized and shifted vertically for clarity. The top spectrum in
each panel corresponds to a velocity standard star observed with the
same instrumental setup.}
\label{fig:Atlas_ESO_a}
\end{figure*}

\begin{figure*}
\includegraphics[width=18cm]{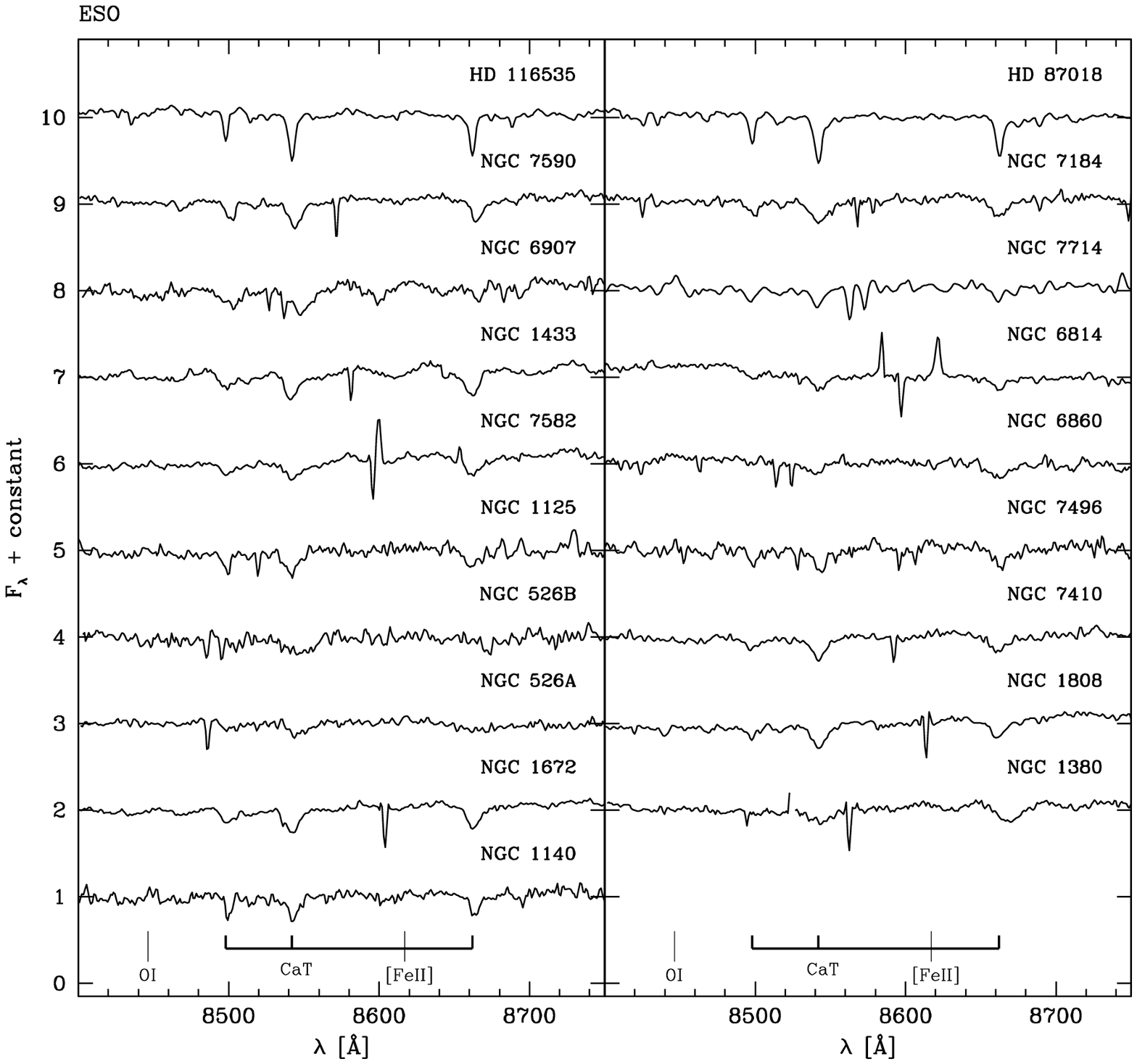}
\caption{As Fig.~\ref{fig:Atlas_ESO_a}.}
\label{fig:Atlas_ESO_b}
\end{figure*}

\begin{figure*}
\includegraphics[width=18cm]{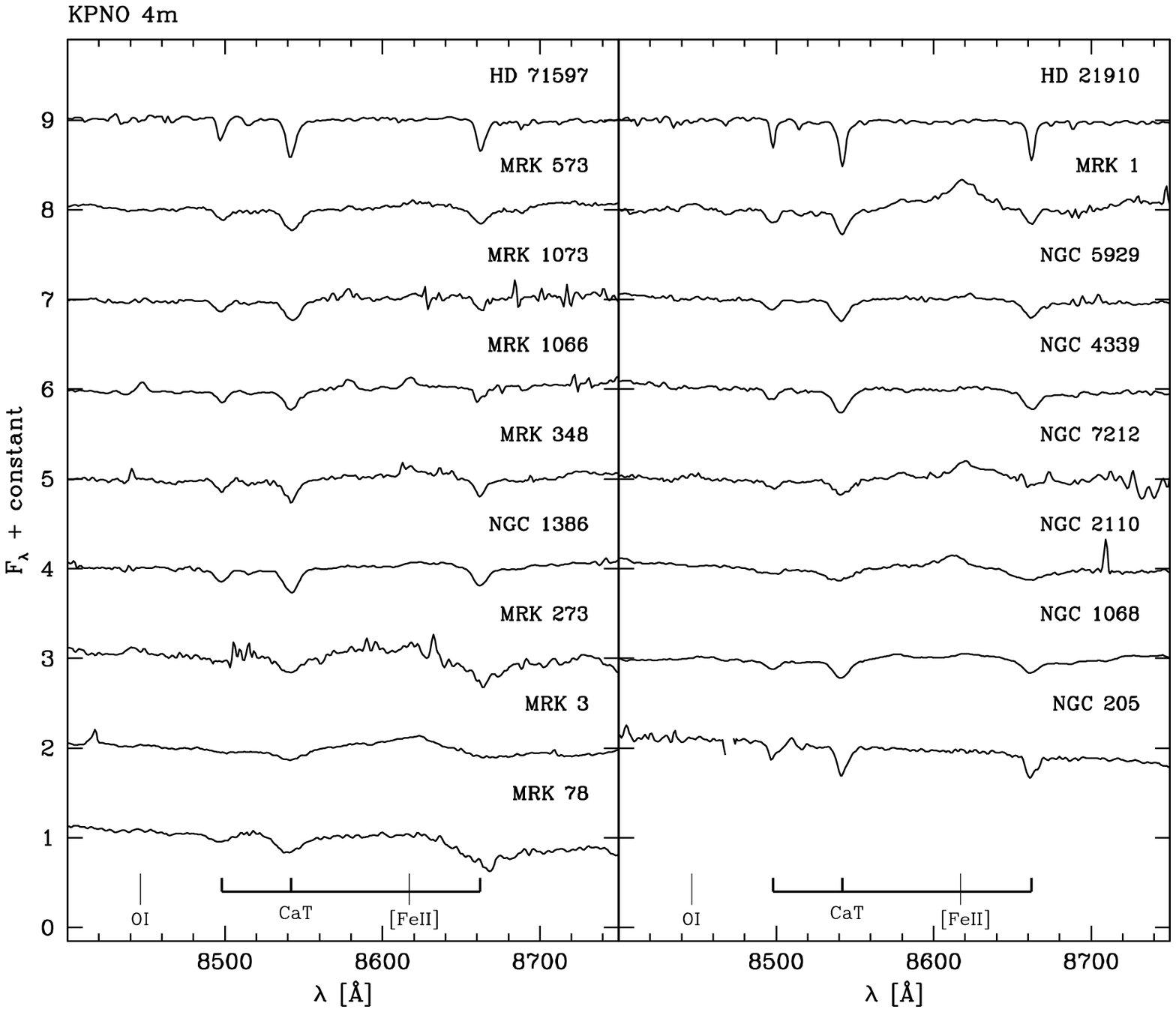}
\caption{As Fig.~\ref{fig:Atlas_ESO_a}, but for observations in the
KPNO 4m telescope.}
\label{fig:Atlas_KPNO4m}
\end{figure*}

\begin{figure*}
\includegraphics[width=18cm]{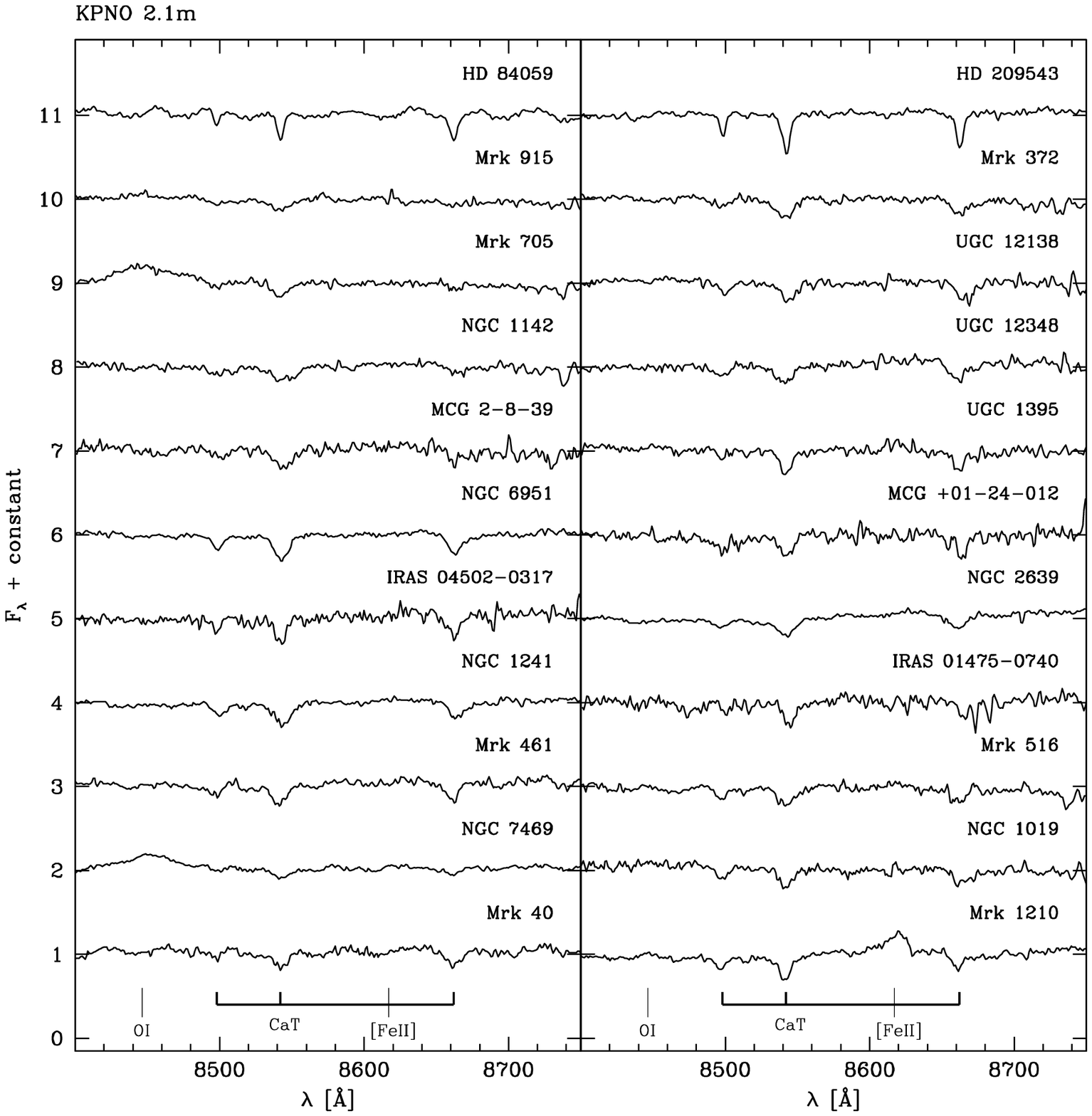}
\caption{As Fig.~\ref{fig:Atlas_ESO_a}, but for observations in the
KPNO 2.1m telescope.}
\label{fig:Atlas_KPNO2m}
\end{figure*}

\begin{figure}
\includegraphics[bb= 50 50 610 610,width=17cm]{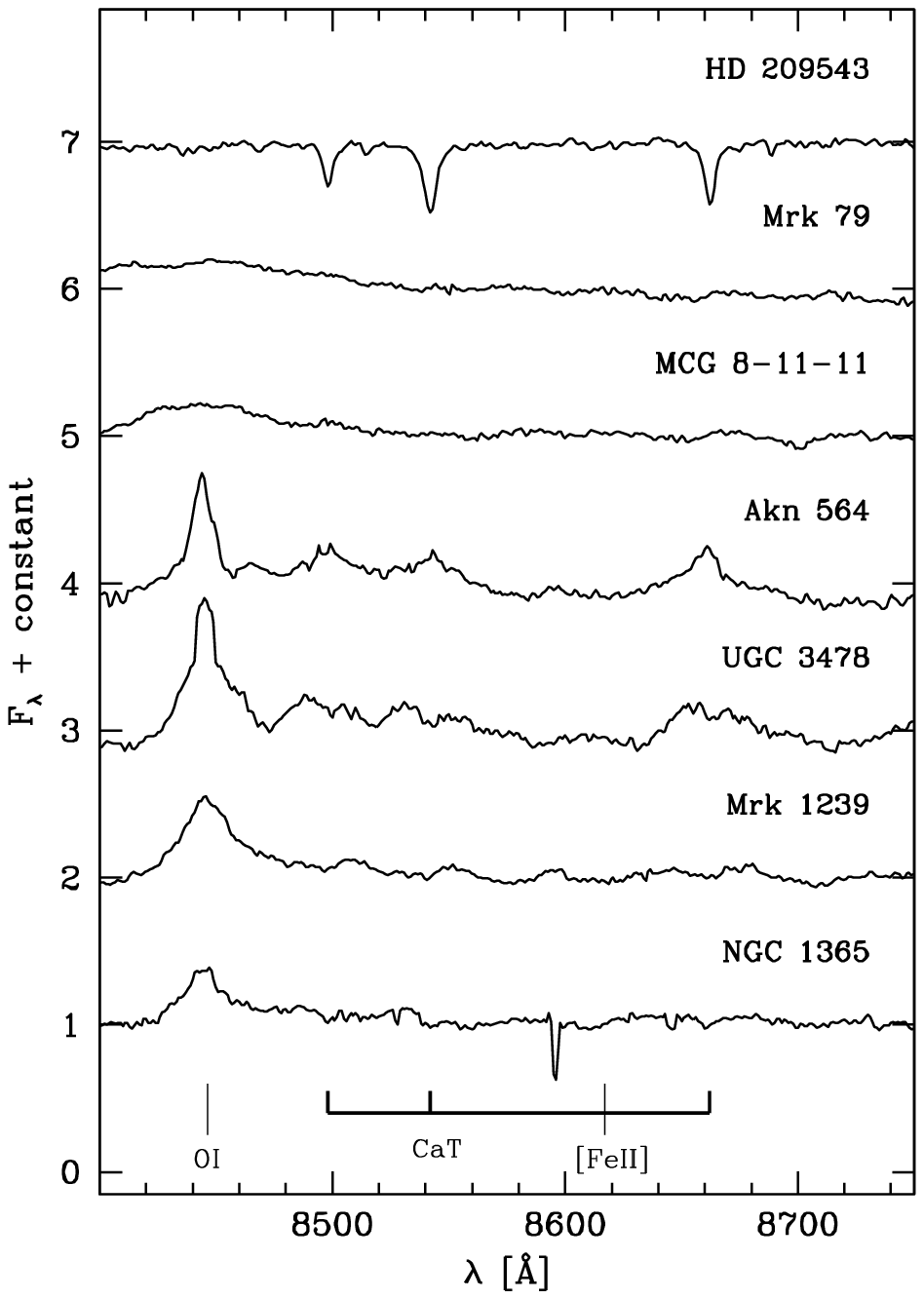}
\vskip -3cm
\caption{Objects containing complex (quality flag = {\it d}) spectra
in the CaT region.}
\label{fig:Atlas_Quality_d}
\end{figure}

\subsection{CaT spectra}

\label{sec:CaT_atlas}

Figures \ref{fig:Atlas_ESO_a}--\ref{fig:Atlas_Quality_d} present the
spectra around the CaT region for all 78 galaxies in our sample.  Data
from different telescopes are displayed in different plots.  As in NW,
each panel in these figures includes an example of a velocity standard
star observed through the same setup, to illustrate the kinematical
broadening in the galaxy spectra.  The wavelengths of the CaT lines
and other features are indicated. We recall that all spectra in this
atlas correspond to the nuclear regions: $2^{\prime\prime} \times
2.46^{\prime\prime}$ for the ESO galaxies, $2^{\prime\prime} \times
2.34^{\prime\prime}$ for the ones observed at the KPNO 2.1m and
$1.5^{\prime\prime} \times 2^{\prime\prime}$ for those observed at the
KPNO 4m.  Defining $r_{ap}$ as the radius of a circle whose area
equals the aperture area, our nuclear spectra corresponds to physical
regions of projected radii between $r_{ap} = 50$--700 pc of the
nucleus, with a median value of 286 pc (Table \ref{tab:sample}).

All spectra were brought to the rest-frame using, whenever possible,
the mean redshift derived from the CaT absorption lines. For
convenience, the spectra were normalized to the median flux in the
8554--8574 \AA\ interval. We estimate the signal-to-noise ratio
($S/N$) by the mean/rms flux ratio in this same window. The resulting
S/N spans the 8--125 range, with a median $S/N$ of 41.

Figures \ref{fig:Atlas_ESO_a}--\ref{fig:Atlas_Quality_d} show that our
NIR spectra often contain features which complicate the analysis of
the CaT, such as emission lines (narrow and broad), noise and
imperfectly removed atmospheric features. To help dealing with this
problem we have assigned a ``quality flag'' to each spectrum according
to the degree of contamination of the CaT lines. Column 8 of Tables
\ref{tab:log_ESO}--\ref{tab:log_ROSA} list the results.  Quality {\it
a} refers to the best spectra, where the CaT lines are little or not
affected by any of the problems above, as in NGC 2997, Mrk 573 and NGC
2639 (Figures \ref{fig:Atlas_ESO_a}, \ref{fig:Atlas_KPNO4m} and
\ref{fig:Atlas_KPNO2m} respectively). Quality {\it b} refers to
reasonably good spectra, but where one of the CaT lines is
contaminated. NGC 3281 (Fig \ref{fig:Atlas_ESO_a}) and Mrk 3 (Fig
\ref{fig:Atlas_KPNO4m}) are examples of quality {\it b}
spectra. Quality {\it c} corresponds to problematic spectra, like NGC
4748 and IRAS 01475-0740 in Figs \ref{fig:Atlas_ESO_a} and
\ref{fig:Atlas_KPNO2m} respectively. CaT measurements for these
objects should be treated with caution.  Finally, we define as quality
{\it d} those spectra which are so complex that it is impossible to
derive any reasonably accurate CaT measurement. All such cases are
presented in Fig.~\ref{fig:Atlas_Quality_d}. Some Seyfert 1s and most
Narrow Line Seyfert 1s in our sample fall in this category. In this
latter class the CaT absorption lines are superposed to broad emission
components of the same transitions from the Broad Line Region (Ferland
\& Persson 1989).  AKN 564 and Mrk 1239 are two such cases. We have not
attempted to disentangle the absorption and emission components in
these cases. In some of these objects useful CaT information can be
obtained from off-nuclear extractions (Paper II).

The above classification allows us to investigate how our results are
affected by data quality. As expected, the $S/N$ ratio tends to
increase as one goes from quality {\it c} to {\it a}. The median
values are $S/N = 26$, 37 and 47 for qualities {\it c}, {\it b} and
{\it a} respectively. Also, the uncertainties in the CaT products
(velocity dispersion and equivalent width) are smaller the better the
quality (Sections \ref{sec:vel_disp} and \ref{sec:EWs}).

Our data comprises 40, 15 and 17 objects with quality flags {\it a},
{\it b} and {\it c} respectively, totaling 72 galaxies with useful CaT
spectra. This statistically significant dataset, which is similar in
quality and quantity to that of NW, is analyzed in the next sections.

\subsection{The [SIII]$\lambda$9069 line}

\label{sec:SIII_atlas}

\begin{figure*}
\includegraphics[width=18cm]{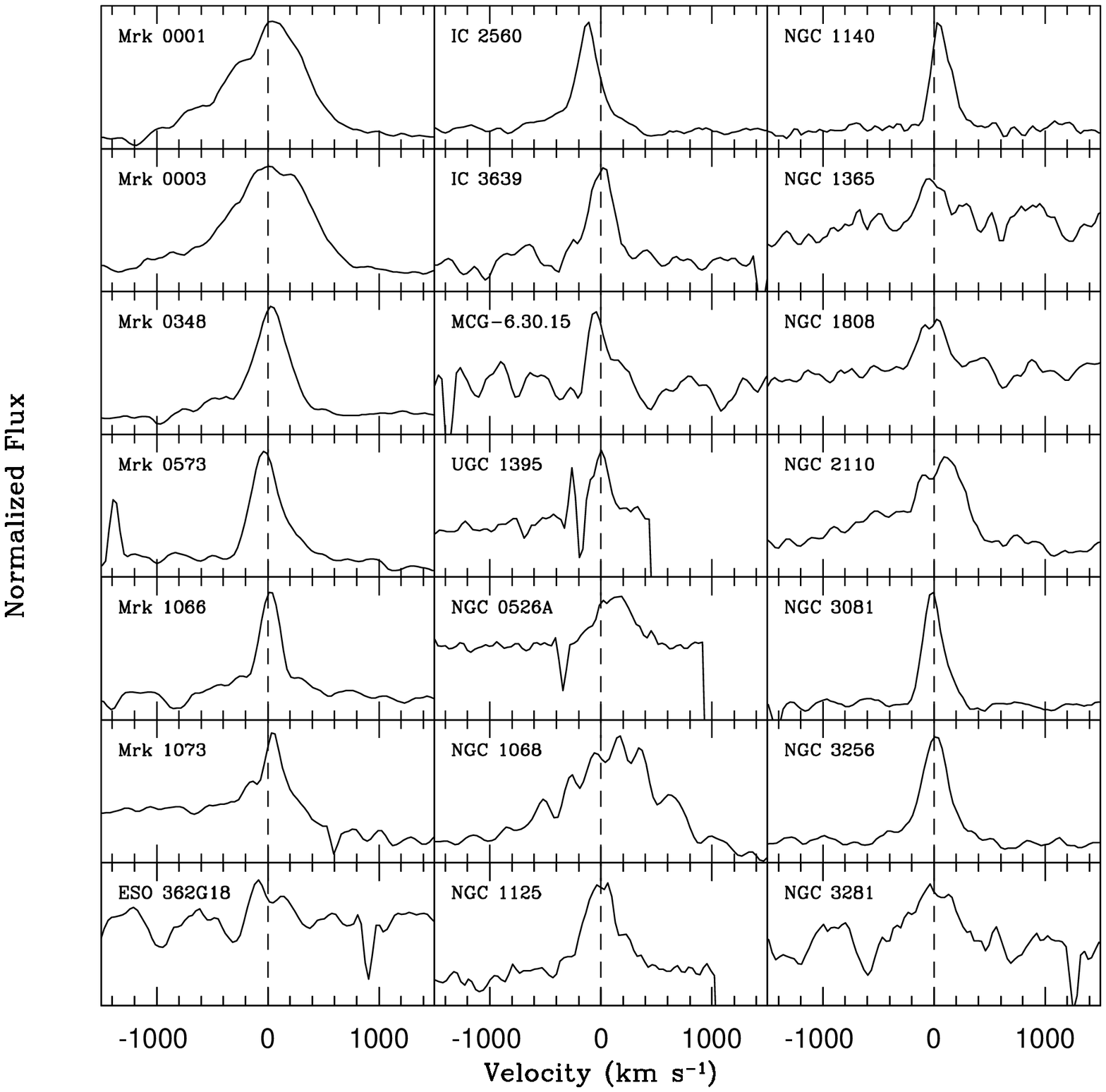}
\caption{[SIII]$\lambda$9069 emission line profiles, normalized to the
continuum level.}
\label{fig:Atlas_SIII_A}
\end{figure*}

\begin{figure*}
\includegraphics[width=18cm]{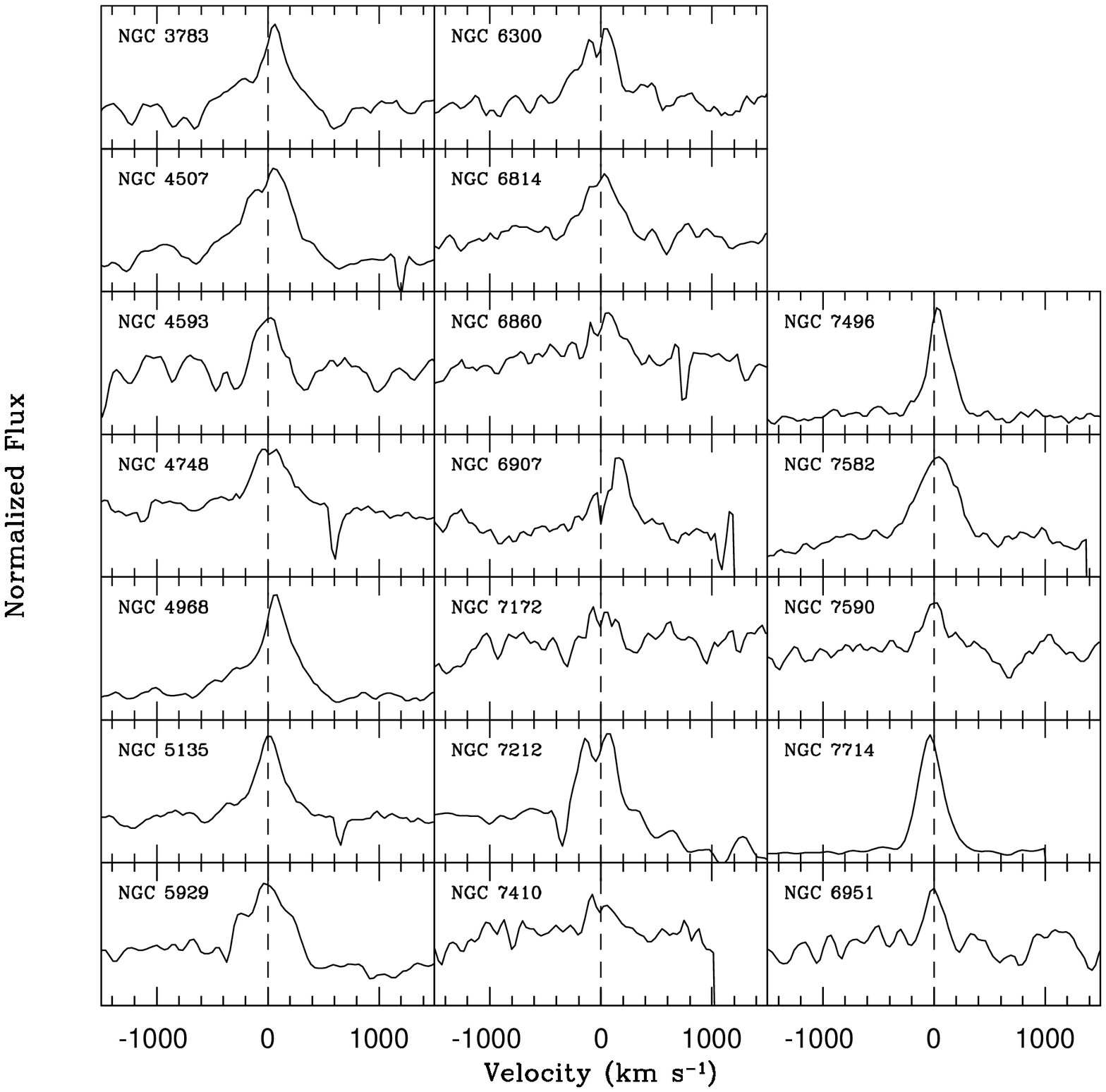}
\caption{As Fig.~\ref{fig:Atlas_SIII_A}.}
\label{fig:Atlas_SIII_B}
\end{figure*}

In 40 galaxies the [SIII]$\lambda$9069 line is detected with at least
a reasonable quality line profile (equivalent width $\ga 1$ \AA). The
[SIII] profiles are shown in Figs~\ref{fig:Atlas_SIII_A} and
\ref{fig:Atlas_SIII_B}. We have fitted this line with a single
Gaussian profile. The resulting values of its width, $\sigma_{\rm
[SIII]}$ (corrected for instrumental broadening) and equivalent width
$W_{\rm [SIII]}$ are listed in Table \ref{tab:results}. We do not
quote [SIII] fluxes because of the uncertain absolute flux scale.  The
width of the [SIII] line provides a rough measure of the typical
velocity of clouds in the Narrow Line Region (NLR) of AGN.  In Paper
II we compare this velocity with the typical stellar velocities
deduced from the analysis of the CaT lines.  Paper II also compares
the $W_{\rm [SIII]}$ values of type 1 and 2 Seyferts, which may
indicate the diluting effects of an underlying AGN continuum at
near-IR wavelengths.

\section{Stellar Velocity Dispersions}

\label{sec:vel_disp}

Due to its location in a relatively clean spectral region, the CaT is
an ideal tracer of stellar kinematics in galaxies. This potential was
recognized long ago in studies of normal galaxies (Pritchet 1978;
Dressler 1984) and even AGN (TDT). This is all the more true in AGN,
were optical kinematical tracers like the Mg lines at $\sim 5175$ \AA~
are often contaminated by emission features, which complicates the
measurement of kinematical properties (NW).  The combination of this
practical advantage with the discovery of the $M_{\rm
BH}$-$\sigma_\star$ relation brought renewed interest in the CaT as a
tool to measure $\sigma_\star$, and thus indirectly weigh black
holes, particularly in AGN.

In this section we present measurements of $\sigma_\star$ for galaxies
in our atlas. Two different methods were employed to estimate
$\sigma_\star$. In what follows we describe these methods
(\S\ref{sec:DFM} and \ref{sec:XCOR}) and compare their results both
between them and with values in the literature
(\S\ref{sec:vd_comparison}).

\subsection{Direct Fitting Method}

\label{sec:DFM}

Kinematical parameters can be estimated by a Direct Fitting Method
(DFM), which consists of fitting a model to the observed CaT spectrum
directly in $\lambda$-space (Barth, Ho \& Sargent 2002).  A model
spectrum $M_\lambda$ can be built combining one or more stellar
templates with a continuum $C_\lambda$, and then convolving it with an
assumed Gaussian line of sight velocity distribution function
$G(v_\star,\sigma_\star)$, centered at $v_\star$ and broadened by
$\sigma_\star$. The resulting expression for $M_\lambda$ is

\begin{equation}
\label{eq:DFM_spectrum}
M_\lambda = 
   M_{\lambda_0}
   \left[
   \sum_{j=1}^{N} x_j T_{j,\lambda} r_\lambda
   \right] 
   \otimes G(v_\star,\sigma_\star)
\end{equation}

\ni where 

\begin{itemize}

\item[(i)] $T_{\lambda,j}$ is the spectrum of the $j^{\rm th}$
template star normalized at $\lambda_0$. The continuum $C_\lambda$ is
also included in the $T_{\lambda,j}$ base as a set of power-laws with
different slopes. Each galaxy was modeled with a base containing only
velocity standard stars observed under the same instrumental setup,
thus circumventing the need for corrections due to different spectral
resolutions.

\item[(ii)] $\vec{x}$ is a vector whose components $x_j$ ($j = 1\ldots
N$) represent the fractional contribution of each base element to the
total synthetic flux at $\lambda_0$, denoted by $M_{\lambda_0}$.

\item[(iii)] $r_\lambda \equiv 10^{-0.4 (A_\lambda - A_{\lambda_0})}$
accounts for reddening by a foreground dust screen.

\end{itemize}

The continuum components in the $T_{\lambda,j}$ base are introduced
with the specific aim of allowing the fits to account for a possible
mismatch between the velocity standard templates and the stars in the
galaxy or dilution of the CaT by an underlying continuum. It is clear
that we cannot hope to constrain well the shape of $C_\lambda$ given
the narrow spectral range of our data. Similarly, extinction is
included in the fits just for completeness, since it cannot be well
constrained by the data either. Indeed, of all $N_\star + 3$
parameters in the model we are only interested in one:
$\sigma_\star$. The effect of $\sigma_\star$ upon $M_\lambda$ is
completely different from that of all the uninteresting parameters,
which allows it to be well constrained by the data.

The fits are performed by minimizing the $\chi^2$ between model and
observed ($O_\lambda$) spectra:

\begin{equation}
\label{eq:DFM_chi2}
\chi^2 =
   \sum_{\lambda}
   \left[
   \left(O_\lambda - M_\lambda \right) w_\lambda
   \right]^2
\end{equation}

\ni where the weight spectrum $w_\lambda$ is defined as the inverse of
the noise in $O_\lambda$. Emission lines and spurious features, such
as residual sky lines, are masked out by setting $w_\lambda =
0$. Except for the spectral base and wavelength range, this method is
{\it identical} to the one in Cid Fernandes \etal (2004, 2005), who
fit the optical spectra of galaxies by a combination of evolutionary
synthesis models with the code STARLIGHT. In fact, exactly the same
code was used in our fits. We thus refer the reader to those papers
for details on the numerical aspects of the fits.

\begin{figure*}
\includegraphics[width=18cm]{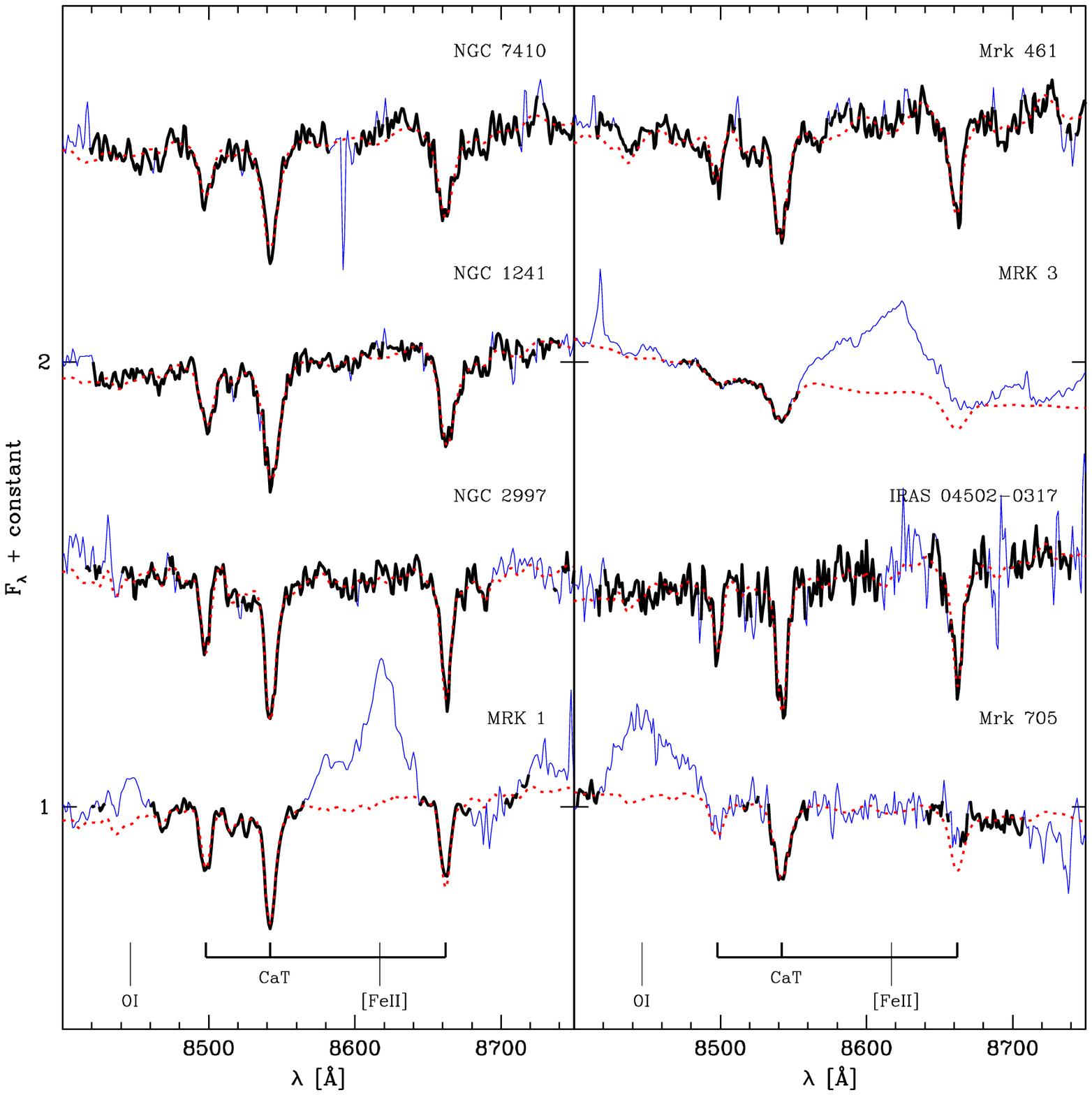}
\caption{Examples of fits to the galaxy spectra. The solid line shows
the observed spectrum, while the dotted line shows the spectral fits
obtained with the DFM. A thicker line is used to mark the region
actually used in the fits. NGC 7410, NGC 1241, NGC 2997 and Mrk 1 have
quality flag {\it a}, Mrk 461 and Mrk 3 have quality {\it b}, IRAS
04502-0317 and Mrk 705 have quality {\it c}.}
\label{fig:DFM_fits}
\end{figure*}

Figure \ref{fig:DFM_fits} illustrates some of the resulting fits. In
objects with clean CaT spectra (like NGC 1241 and NGC 2997), spectral
regions outside the CaT lines were included in the fit, but we
emphasize that ignoring them yields practically identical values of
$\sigma_\star$. For more problematic spectra (e.g., Mrk 3 and Mrk
705), we have chosen to concentrate the fits on the CaT lines, masking
out other wavelengths.  The exact choice of the mask affects the
derived values of $\sigma_\star$, but these variations are well within
the uncertainties $\Delta \sigma_\star$ estimated below. Table
\ref{tab:results} lists the resulting values of $\sigma_\star$.

The uncertainty in $\sigma_\star$ was estimated by the method outlined
in Barth \etal (2002), which basically consists of finding the range
in $\sigma_\star$ which causes an increase of $\Delta \chi^2 = 1$ over
the best fit, after rescaling the errors to yield a best fit reduced
$\chi^2$ of 1. We find $\Delta \sigma_\star$ in the 3--24  km/s
range (Table \ref{tab:results}). Separating spectra according to their
quality flags, we find median values of $\Delta \sigma_\star = 6$,
10 and 12 km/s for quality {\it a}, {\it b} and {\it c}
respectively.

An independent assessment of the uncertainties is possible in the
cases of Mrk 1210 and NGC 7130, for which we have repeated
observations with different telescopes. For Mrk 1210, we obtain
$\sigma_\star = 72 \pm 9$ and $77 \pm 7$ km/s for the ESO and KPNO 2.1
observations, respectively, while for NGC 7130 we derive $\sigma_\star
= 140 \pm 8$ and $112 \pm 9$ km/s for the ESO and KPNO 4m data. In
both cases the independent measurements are consistent to within $\sim
2$ sigma confidence.

A more traditional method to estimate $\Delta \sigma_\star$ consists
of evaluating the dispersion among fits performed using individual
template stars (e.g., Tonry \& Davis 1979). For completeness, we have
also evaluated $\Delta \sigma_\star$ with this method. The
uncertainties obtained in this way are $\sim 30$ per cent larger than
those obtained by the method described above.  This agreement is
hardly surprising, given that template mismatch is already accounted
for in our implementation of the DFM, which includes a library of
template stars in the base $T_{j,\lambda}$ (equation
\ref{eq:DFM_spectrum}). When performing fits with a fixed
$\sigma_\star$, as we do in the estimation of $\Delta \sigma_\star$
following the methodology of Barth \etal (2002), the fractions $x_j$
associated with each template star are allowed to vary freely, which
is qualitatively equivalent to changing the template star.  Hence,
unlike in Barth \etal (2002), there is no need to add these two
estimates of $\Delta \sigma_\star$ in quadrature to obtain a total
error estimate. (Doing that would increase $\Delta \sigma_\star$ by
$\sim 60$ percent with respect to the values listed in Table
\ref{tab:results}.)

\begin{table*}
\small
\begin{center}
\begin{tabular}{lrrrrr}
\hline
Object                      & 
$\sigma_\star^{\rm DFM}$    &  
$\sigma_\star^{\rm CCM}$   &
$\sigma_{\rm [SIII]}$       & 
$W_{\rm [SIII]}$            & 
$W_{\rm CaT}$               \\
           &
(km/s)     &
(km/s)     &
(km/s)     &
(\AA)      &
(\AA)      \\ \hline
ESO 362-G08	&   179 $\pm$  7 	&    193 $\pm$ 10 	&	   	&	   	&	7.2 $\pm$ 0.3 \\
ESO 362-G18	&   126 $\pm$  5 	&    134 $\pm$  4 	&	132:	&	  3:	&	6.7 $\pm$ 0.3 \\
IC  2560	&   135 $\pm$  4 	&    138 $\pm$  5 	&	107	&	 16	&	7.8 $\pm$ 0.4 \\
IC  3639	&    95 $\pm$  5	&     99 $\pm$  5 	&	111	&	 10	&	6.3 $\pm$ 0.3 \\
IC  5169	&   114 $\pm$ 12 	&    111 $\pm$  2 	&	   	&	   	&	7.5 $\pm$ 0.9 \\
IRAS 01475-0740	&    62 $\pm$ 11 	&    108 $\pm$ 17 	&	   	&	   	&	6.0 $\pm$ 0.3 \\
IRAS 04502-0317	&    74 $\pm$  8	&     74 $\pm$ 15 	&	   	&	   	&	6.8 $\pm$ 0.5 \\
MCG +01-24-012	&    84 $\pm$ 10 	&     92 $\pm$ 18 	&	   	&	   	&	6.8 $\pm$ 0.7 \\
MCG -6.30.15	&    94 $\pm$  8 	&    103 $\pm$  4 	&	 90	&	  5	&	5.1 $\pm$ 0.5 \\
MCG 2-8-39	&   170 $\pm$ 13 	&    126 $\pm$ 11 	&	   	&	   	&	8.0 $\pm$ 0.8 \\
MRK    1	&    86	$\pm$ 4 	&     79 $\pm$  4 	&	375	&	 86	&	6.2 $\pm$ 0.5 \\
MRK    3	&   228 $\pm$ 13 	&    249 $\pm$  4 	&	364	&	 42	&	4.2 $\pm$ 0.4 \\
MRK   78	&   201 $\pm$  8 	&    186 $\pm$  4 	&	   	&		&	7.3 $\pm$ 0.3 \\
MRK  273	&   211 $\pm$ 14 	&    186 $\pm$  2 	&	   	&	   	&	7.7 $\pm$ 0.7 \\
MRK  348	&    95 $\pm$  6 	&     98 $\pm$  8 	&	179	&	 40	&	6.3 $\pm$ 0.2 \\
MRK  573	&   147 $\pm$  5 	&    148 $\pm$  3 	&	171	&	 28	&	7.9 $\pm$ 0.1 \\
MRK 1066	&   100 $\pm$  4 	&     90 $\pm$  6 	&	119	&	 21	&	5.9 $\pm$ 0.4 \\
MRK 1073	&   114 $\pm$  6 	&    109 $\pm$  5 	&	168	&	 24	&	5.9 $\pm$ 0.5 \\
Mrk   40	&   125 $\pm$  7 	&    116 $\pm$  4 	&	   	&	   	&	4.0 $\pm$ 0.4 \\
Mrk  372	&   155 $\pm$  6 	&    161 $\pm$  5 	&	   	&	   	&	6.4 $\pm$ 0.5 \\
Mrk  461	&   111 $\pm$  6 	&    123 $\pm$  4 	&	   	&	   	&	5.6 $\pm$ 0.4 \\
Mrk  516	&   113 $\pm$ 12 	&    114 $\pm$  7 	&	   	&	   	&	7.4 $\pm$ 0.6 \\
Mrk  705	&   128 $\pm$ 11 	&    120 $\pm$ 15 	&	   	&	   	&	5.1 $\pm$ 0.3 \\
Mrk  915	&   181 $\pm$ 18 	&    146 $\pm$ 16 	&	   	&	   	&	5.2 $\pm$ 0.6 \\
Mrk 1210	&    77 $\pm$  7 	&     82 $\pm$ 16 	&	   	&	   	&	6.7 $\pm$ 0.4 \\
NGC  205	&    47 $\pm$  6 	&     74 $\pm$  6 	&	   	&	   	&	6.2 $\pm$ 0.2 \\
NGC  526A	&   198 $\pm$ 16 	&    219 $\pm$ 11 	&	159	&	  4	&	4.7 $\pm$ 0.5 \\
NGC  526B	&   237 $\pm$ 22 	&    167 $\pm$ 11 	&	   	&	   	&	7.0 $\pm$ 1.1 \\
NGC 1019	&   106 $\pm$  9 	&    110 $\pm$ 11 	&	   	&	   	&	6.5 $\pm$ 0.5 \\
NGC 1068	&   140 $\pm$  6 	&    147 $\pm$  3 	&	543	&	120	&	6.3 $\pm$ 0.4 \\
NGC 1125	&   118 $\pm$  9 	&    138 $\pm$  6 	&	168	&	 11	&	7.6 $\pm$ 0.5 \\
NGC 1140	&    53 $\pm$  6 	&     60 $\pm$  3 	&	 81	&	 12	&	5.9 $\pm$ 0.7 \\
NGC 1142	&   219 $\pm$ 15 	&    202 $\pm$ 47 	&	   	&	   	&	8.6 $\pm$ 0.4 \\
NGC 1241	&   136 $\pm$  5 	&    142 $\pm$ 12 	&	   	&	   	&	8.5 $\pm$ 0.4 \\
NGC 1380	&   250 $\pm$ 16 	&    215 $\pm$  8 	&	   	&	   	&	7.8 $\pm$ 2.4 \\
NGC 1386	&   123 $\pm$  3 	&    133 $\pm$  3 	&	   	&	   	&	8.1 $\pm$ 0.2 \\
NGC 1433	&    98 $\pm$  6 	&    113 $\pm$  3 	&	   	&	   	&	7.6 $\pm$ 0.4 \\
NGC 1672	&   108 $\pm$  4 	&    111 $\pm$  3 	&	   	&	   	&	7.7 $\pm$ 0.2 \\
NGC 1808	&   119 $\pm$  6 	&    129 $\pm$  4 	&	129	&	  2	&	7.3 $\pm$ 0.4 \\
NGC 2110	&   264 $\pm$ 11 	&    273 $\pm$  7 	&	375	&	 16	&	6.4 $\pm$ 0.3 \\
NGC 2639	&   168 $\pm$  6 	&    155 $\pm$ 12 	&	   	&	   	&	7.2 $\pm$ 0.2 \\
NGC 2997	&    79 $\pm$  4	&     89 $\pm$  4 	&	   	&	   	&	8.0 $\pm$ 0.4 \\
NGC 3081	&   129 $\pm$  8 	&    113 $\pm$  4 	&	 77	&	 20	&	7.4 $\pm$ 0.4 \\
NGC 3115	&   275 $\pm$  6 	&    268 $\pm$  8 	&	   	&	   	&	7.0 $\pm$ 0.4 \\
NGC 3256	&   130 $\pm$ 13 	&    100 $\pm$  6 	&	120	&	 17	&	3.7 $\pm$ 0.5 \\
NGC 3281	&   161 $\pm$  8 	&    176 $\pm$  3 	&	235	&	  8	&	7.3 $\pm$ 0.4 \\
NGC 3783	&   116 $\pm$ 20 	&    114 $\pm$  6 	&	247	&	 17	&	3.0 $\pm$ 0.2 \\
NGC 4339	&   123 $\pm$  3 	&    129 $\pm$  3 	&	   	&	   	&	7.4 $\pm$ 0.2 \\
NGC 4507	&   146 $\pm$  7 	&    152 $\pm$  4 	&	229	&	 16	&	7.0 $\pm$ 0.4 \\
NGC 4593	&   153 $\pm$ 24 	&    105 $\pm$  5 	&	 96	&	  3	&	3.4 $\pm$ 0.3 \\
NGC 4748	&    76 $\pm$ 15 	&     78 $\pm$ 13 	&	187	&	 10	&	3.4 $\pm$ 0.5 \\
NGC 4968	&   105 $\pm$  9 	&    106 $\pm$  4 	&	182	&	 16	&	6.9 $\pm$ 0.5 \\
NGC 5135	&   128 $\pm$  8 	&    124 $\pm$  6 	&	135	&	 10	&	6.1 $\pm$ 0.4 \\
NGC 5929	&   119 $\pm$  4 	&    122 $\pm$  4 	&	195	&	  9	&	6.5 $\pm$ 0.2 \\
NGC 6300	&    92 $\pm$  5 	&    110 $\pm$  5 	&	217	&	  7	&	8.3 $\pm$ 0.4 \\
NGC 6814	&    83 $\pm$ 11 	&    113 $\pm$  6 	&	169	&	  5	&	4.0 $\pm$ 0.3 \\
NGC 6860	&   162 $\pm$ 11 	&    141 $\pm$  5 	&	153:	&	  3:	&	5.6 $\pm$ 0.6 \\
NGC 6907	&   157 $\pm$ 12 	&    195 $\pm$ 15 	&	199	&	  6	&	9.2 $\pm$ 1.0 \\
NGC 6951	&   115 $\pm$  4 	&    113 $\pm$ 12 	&	 73	&	  2	&	9.0 $\pm$ 0.3 \\ \hline
\end{tabular} 
\end{center}
\caption{Columns 2 and 3: Velocity dispersions obtained with the DFM
and CCM methods respectively. Columns 4 and 5: Width and equivalent
width of the [SIII]$\lambda$9069 emission line. Uncertain measurements
are marked with a `:'.  Column 6: CaT equivalent width.}
\label{tab:results}
\end{table*}

\begin{table*}
\small
\begin{center}
\begin{tabular}{lrrrrr}
\hline
Object                      & 
$\sigma_\star^{\rm DFM}$    &  
$\sigma_\star^{\rm CCM}$   &
$\sigma_{\rm [SIII]}$       & 
$W_{\rm [SIII]}$            & 
$W_{\rm CaT}$               \\
           &
(km/s)     &
(km/s)     &
(km/s)     &
(\AA)      &
(\AA)      \\ \hline
NGC 7130	&   141 $\pm$  8 	&    147 $\pm$  5 	&	   	&	   	&	6.9 $\pm$ 0.4 \\
NGC 7172	&   154 $\pm$  6 	&    160 $\pm$  9 	&	112:	&	  1:	&	6.9 $\pm$ 1.1 \\
NGC 7184	&   146 $\pm$  7 	&    131 $\pm$  5 	&	   	&	   	&	7.9 $\pm$ 1.2 \\
NGC 7212	&   143 $\pm$ 10 	&    140 $\pm$  2 	&	164	&	 26	&	5.3 $\pm$ 0.2 \\
NGC 7410	&   144 $\pm$  7 	&    144 $\pm$  6 	&	126:	&	  1:	&	7.7 $\pm$ 0.4 \\
NGC 7469	&   125 $\pm$ 12 	&    144 $\pm$ 11 	&	   	&	   	&	2.9 $\pm$ 0.2 \\
NGC 7496	&    76 $\pm$ 10	&     94 $\pm$  6 	&	 96	&	 15	&	5.8 $\pm$ 0.7 \\
NGC 7582	&   121 $\pm$  7 	&    113 $\pm$  3 	&	180	&	  8	&	5.9 $\pm$ 0.5 \\
NGC 7590	&    93 $\pm$  4	&     90 $\pm$  4 	&	136	&	  2	&	7.7 $\pm$ 2.0 \\
NGC 7714	&    59 $\pm$  9 	&     65 $\pm$  4 	&	105	&	 28	&	4.7 $\pm$ 2.4 \\
UGC  1395	&    66 $\pm$  6 	&     62 $\pm$ 16 	&	 47:	&	  4:	&	6.5 $\pm$ 0.4 \\
UGC 12138	&   115 $\pm$ 10 	&    136 $\pm$  8 	&	   	&	   	&	6.9 $\pm$ 0.6 \\
UGC 12348	&   155 $\pm$  9 	&    165 $\pm$ 14 	&	   	&	   	&	7.6 $\pm$ 0.4 \\ \hline
\end{tabular} 
\end{center}
\caption{Continuation of Table \ref{tab:results}.}
\end{table*}

\subsection{Cross-Correlation Method}

\label{sec:XCOR}

One of the first techniques devised to measure velocity dispersions in
galaxies is the cross-correlation method (CCM, Tonry \& Davis
1979). NW review this method and show that in the CaT region it yields
fairly good results. In this work, the IRAF task {\it fxcor} was
used. In few words, {\it fxcor} finds the cross-correlation function
between the galaxy and the template spectra in Fourier space. The peak
of this function is then modeled by a Gaussian. We use the same
individual masks which were used in the DFM, and we allow a linear
continuum subtraction from the galaxy spectra. The output from {\it
fxcor} is calibrated in order to account for instrumental resolution.
Column 3 of Table \ref{tab:results} presents $\sigma_\star$ values
obtained with this method. Uncertainties in this case are evaluated
from the rms in $\sigma_\star$ values obtained using different
template stars.

We find that the DFM and the CCM yield velocity dispersions consistent
to within 19 km/s on-average. The agreement is much better for quality
{\it a} spectra, for which the difference between $\sigma_\star^{\rm
DFM}$ and $\sigma_\star^{\rm CCM}$ is just 9 km/s in the rms. For
quality {\it b} and {\it c} the DFM and CCM methods agree to rms
dispersions of 20 and 30 km/s respectively, which further confirms
that data quality is the major source of uncertainty in
$\sigma_\star$.  The uncertainties $\Delta \sigma_\star$ obtained with
these two methods are also similar, with an rms difference of 6 km/s.

\subsection{Comparison with previous results}

\label{sec:vd_comparison}

\begin{figure}
\includegraphics[width=9cm]{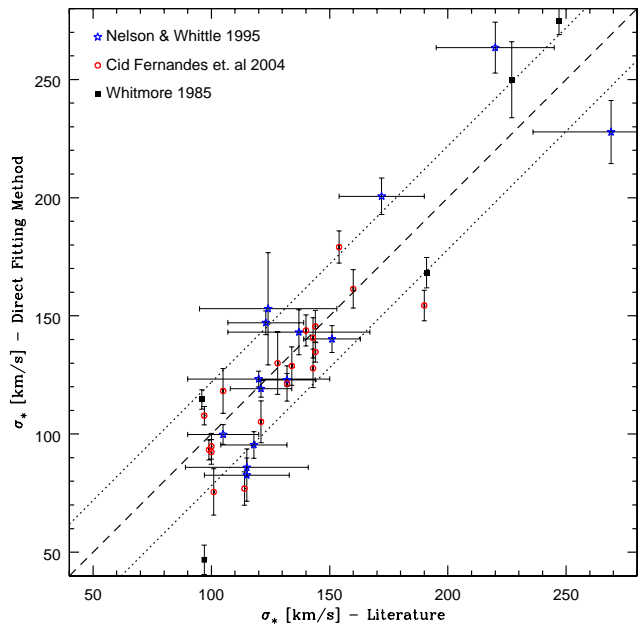}
\caption{Comparison of the stellar velocity dispersions estimated in
this work and values compiled from the literature.  The identity line
is traced by the dashed line, while dotted lines indicate the $\pm 22$
km/s global dispersion ($ = 1$ sigma).}
\label{fig:vd_US_X_Literature}
\end{figure}

In Fig.~\ref{fig:vd_US_X_Literature} we compare our DFM measurements
of $\sigma_\star$ with values compiled from the literature for objects
in common. The two major sources of $\sigma_\star$ are NW and Cid
Fernandes \etal (2004), which have 14 and 22 objects in common
with our sample, respectively. NW derive $\sigma_\star$ from the
cross-correlation method applied to either the CaT or the Mg lines in
the optical, while Cid Fernandes \etal (2004) estimate $\sigma_\star$
fitting the 3500--5200 \AA\ spectra of Seyfert 2s with a combination
of single stellar populations from the Bruzual \& Charlot (2003,
hereafter BC03) models. These two studies agree in their estimates of
$\sigma_\star$ at the level of $\pm 21$ km/s rms.

Our CaT-based estimates of $\sigma_\star$ are in good agreement with
these previous estimates. The rms difference between our values and
those in the literature is 22 km/s.  The spread is somewhat smaller
for quality {\it a} data (rms of 21 km/s) than for qualities {\it b}
and {\it c} (25 km/s). Furthermore, on average, our estimates of
$\sigma_\star$ are just 4 km/s lower than those in the literature.
Very similar results are obtained comparing our CCM-based estimates of
$\sigma_\star$ with literature data, which also yields an rms of 22
km/s. Given the differences in data quality, spatial extractions,
method of analysis, and the intrinsic uncertainties in $\sigma_\star$,
we conclude that there are no significant differences between our
estimates of $\sigma_\star$ and those reported in previous studies.
In any case, the level of agreement is very similar to those among
other studies.

We also made a compilation from Oliva et al. (1995, 1999) near-IR CO and Si $\sigma_\star$ estimates, 
where we can find 9 objects in common with our sample. Although the two estimates are very likely to be 
correlated (P$_{rs}\approx$ 2.5\%), here the difference between results is higher than when compared with optical data, indicating that
our estimated $\sigma_\star$ is, on average, lower by 34 km/s. Considering just the 6 objects that have an "a"
fitting quality, this difference practically does not change (33 km/s), having a rms of 38 km/s (i.e., 
agrees with our results within a 0.05 confidence level, considering that their errors are also of 20 km/s).

\section{Equivalent Widths}

\label{sec:EWs}

Besides its use as a tracer of stellar kinematics, the CaT provides
information on the properties of stellar populations in galaxies. The
most extensive and up-to-date study of the CaT in stars and as a
diagnostic of stellar populations in galaxies has been published by
Cenarro and co-workers in the last few years (Cenarro \etal 2001a, 2001b, 
2002, 2003; Vazdekis \etal 2003). As summarized in the introduction
(see also Paper II), recent work reveals serious difficulties in
interpreting the observed strength of the CaT in normal galaxies,
which in turn raises doubts as to the usefulness of the CaT as a
stellar population diagnostic. Clearly, the situation in AGN ought to
be even more problematic than for normal galaxies, given: (1) the
presence of emission lines around the CaT, (2) the fact that many AGN
are known to be surrounded by starbursts of various ages and
intensities, and (3) the presence of an underlying non-stellar
continuum, seen either directly (in Seyfert 1s) or scattered (Seyfert
2s).

Notwithstanding these caveats, in this section we present measurements
of the CaT equivalent width for objects in our sample. These data is
used in Paper II to evaluate the presence of an AGN continuum at NIR
wavelengths, and to investigate how our AGN fit into the
CaT-$\sigma_\star$ relation (e.g., Michielsen \etal 2003).

We measure the CaT strength following Cenarro \etal (2001a), who offer
two definitions of the CaT equivalent width: `CaT' (which we call
$W_{\rm CaT}$), which consists of a sum of the equivalent widths of
all three CaT lines, and `CaT*' (called $W_{\rm CaT*}$ here), which
corrects $W_{\rm CaT}$ for contamination by Paschen line absorption.
These equivalent widths are measured with respect to a continuum
defined by fitting the spectrum in 5 windows in the 8474--8792 \AA\
range.

It is evident from Figs.~\ref{fig:Atlas_ESO_a}--\ref{fig:Atlas_KPNO2m}
that this recipe cannot be blindly applied to our spectra, given the
presence of emission lines, atmospheric and noise artifacts which
affect both the continuum, CaT and Paschen bandpasses. On the other
hand, in Section \ref{sec:DFM} we have seen that the DFM provides good
matches to the clean regions of the spectra.  Naturally, these model
spectra do not suffer from the aforementioned problems.  Measuring the
CaT indices over the model fits is therefore a simpler alternative to
removing the unwanted features from the observed spectra.  (A similar
strategy was employed by Cid Fernandes \etal 2004 in their analysis of
optical spectra of Seyfert 2s.) We thus opted to measure both $W_{\rm
CaT}$ and $W_{\rm CaT*}$ over the synthetic spectra.

We tested the validity of this procedure by comparing the values of
$W_{\rm CaT}$ measured for the model ($M_\lambda$) and observed
($O_\lambda$) spectra for galaxies with no serious contamination, or
galaxies where the spurious features can be easily removed (say, by
chopping narrow emission lines). We find that these two estimates of
$W_{\rm CaT}$ agree well, with an rms difference of 0.4 \AA. There is
a small bias, in the sense that $W_{\rm CaT}$ values measured over the
synthetic spectra are on average 0.5 \AA\ smaller than those measured
over the observed spectrum, generally due to noise in the redder
continuum band. Both this offset and the rms difference are comparable
to the formal uncertainties in $W_{\rm CaT}$.  We thus conclude that
this experiment validates our strategy of measuring $W_{\rm CaT}$ over
the model spectra.

Although Cenarro \etal (2001a) and subsequent studies of normal
galaxies favour $W_{\rm CaT*}$ over $W_{\rm CaT}$ as a measure of the
CaT strength, we have reasons to do the opposite choice.  Firstly, in
several cases our DFM fits concentrate on windows centered nearly
exclusively on the CaT lines. Hence, Paschen lines, even if present in
the spectrum (which is not the case in any of our spectra, with the
possible exception of the normal galaxy NGC 6907), would not have a
relevant weight in the spectral fits. Secondly, only the KPNO 2.1m
observations include velocity standard stars with Paschen absorption
lines in their spectra (F giants; see Table
\ref{tab:VelocityStandards}). For the other runs, the spectral base
$T_{j,\lambda}$ does not cover such spectral types, so the
synthetic spectrum cannot possibly model any Paschen line absorption
properly, thus rendering the Paschen line correction in $W_{\rm CaT*}$
meaningless.  Hence, we only present $W_{\rm CaT}$ measurements 
(Table~\ref{tab:results}).

Since the $M_\lambda$ spectra are effectively noiseless, in order to
compute uncertainties in $W_{\rm CaT}$ we carried out a formal
propagation using the $S/N$ derived for each galaxy (Section
\ref{sec:CaT_atlas}).  To be on the conservative side, we further add
0.5 \AA\ in quadrature to account for the empirically established rms
difference between $W_{\rm CaT}$ measurements performed over
$M_\lambda$ and over clean $O_\lambda$ spectra. The median
uncertainties are $\Delta W_{\rm CaT} = 0.6$, 0.7 and 0.8 \AA\ for
quality {\it a}, {\it b} and {\it c} spectra respectively.

Inspection of the $W_{\rm CaT}$ values in Table~\ref{tab:results}
shows that this index spans a range of values from $\sim 1$ to 9 \AA.
However, most values are within the 6--8 \AA\ range, giving the false
impression that stellar populations are very homogeneous in our
sample.  This is definitely not true, as we know from independent work
at other wavelengths for many of the same galaxies studied here (e.g.,
Cid Fernandes \etal 2001, 2004). Instead, the small variation in CaT
strength reinforces the notion that this feature is not much sensitive
to stellar population properties. The only noticeable trend is that
Seyfert 1s tend to have weaker CaT than other galaxies. On the mean,
$W_{\rm CaT} = 4.6 \pm 2.0$ \AA\ (sample dispersion) for Seyfert 1s
and $7.0 \pm 1.0$ \AA\ for Seyfert 2s. The statistics for the 3
starbursts in our sample is $4.7 \pm 1.1$ \AA, while for the remaining
6 non-active galaxies $W_{\rm CaT} = 7.7 \pm 1.0$ \AA, very similar to
the values spanned by Seyfert 2s. The most likely origin for the
difference between type 1 and type 2 AGN is dilution by an underlying
non-stellar featureless continuum in type 1s. Hence, if on the one
hand $W_{\rm CaT}$ turns out to be a rather poor tracer of stellar
populations, it seems to be a good FC detector.  These aspects are
explored in greater depth in Paper II.

\section{Conclusions}

\label{sec:Conclusions}

This paper presented a spectroscopic atlas of 78 galaxies in the region
around the Calcium triplet (CaT). Most of the objects are AGN, split
into 43 type 2 Seyferts and 26 type 1s. The spectra cover the inner
$r_{pc} \sim 300$ pc with a typical S/N of 40. Quality flags were
assigned to each spectrum to describe the degree of contamination of
the CaT absortion lines by emission lines or atmospheric features. The
main products from our analysis of this data set are stellar velocity
dispersions, [SIII]$\lambda$9069 line profiles, and CaT equivalent
widths.

Stellar velocity dispersions ($\sigma_\star$) were measured for 72 of
the spectra using both direct fitting and cross-correlation methods.
The two techniques yield results compatible to within an rms of 19
km/s, which is also the typical uncertainty of our estimates.
Comparison with $\sigma_\star$ values reported in the literature for
objects in common agree with our estimates at the level of $\sim 20$
km/s rms for optical data, and at a 2 sigma level for near-IR data.

We have also analyzed the [SIII]$\lambda$9069 line profiles for 40
galaxies in the sample with useful data in this range. Single Gaussian
fits were performed, producing estimates of line width and equivalent
width.

The CaT equivalent width was measured over the synthetic spectra
obtained from the direct fits, circumventing the manual editing of
the spectra which would be required to remove undesirable features
which affect $W_{\rm CaT}$.  We find that the value of $W_{\rm CaT}$
in Seyfert 2s and normal galaxies are concentrated in a relatively
small range, from $\sim 6$ to 8 \AA.  Type 1 Seyferts tend to have a
weaker CaT, most likely due to dilution by a non-stellar continuum.

These data products are analyzed in Paper II, where we investigate the
relation between nebular and stellar kinematics and the behaviour of
the CaT strength as a function of activity type and stellar population
properties.

\section*{ACKNOWLEDGMENTS}

AGR, LRV, NVA, RCF and TSB acknowledge the support from CAPES, CNPq
and Instituto do Mil\^enio. RGD acknowledges support by Spanish
Ministry of Science and Technology (MCYT) through grant
AYA-2004-02703.  AGR and LRV aknowledge A. Rodr\'{\i}guez-Ardila for 
suggestions in the reduction process. We thank the Laborat\'orio Nacional de
Astrof\'{\i}sica for the allocation of time on the ESO 1.52m and
financial support during the runs. Part of the data described here
were taken at Kitt Peak National Observatory, National Optical
Astronomy Observatories, which are operated by AURA, Inc., under a
cooperative agreement with the National Science Foundation. 
Basic research at the US Naval Research Laboratory is supported by
the Office of Naval Research.
This research made use of the NASA/IPAC
Extragalactic Database (NED), which is operated by the Jet Propulsion
Laboratory, Caltech, under contract with NASA.

\end{document}